\documentclass[fleqn,usenatbib]{mnras}

\usepackage{newtxtext,newtxmath}

\usepackage[T1]{fontenc}

\DeclareRobustCommand{\VAN}[3]{#2}
\let\VANthebibliography\thebibliography
\def\thebibliography{\DeclareRobustCommand{\VAN}[3]{##3}\VANthebibliography}

\usepackage{graphicx}	
\usepackage{amsmath}	

\title[The IMF and multiplicity of stars]{The IMF and multiplicity of stars from gravity, turbulence, magnetic fields, radiation and outflow feedback}

\author[Sajay Sunny Mathew, Christoph Federrath]{
Sajay Sunny Mathew,$^{1,2}$\thanks{E-mail: \href{mailto:sajaysmathew@gmail.com}{sajaysmathew@gmail.com}}
Christoph Federrath$^{1}$\thanks{E-mail: \href{mailto:christoph.federrath@anu.edu.au}{christoph.federrath@anu.edu.au}}
\\
$^{1}$Research School of Astronomy and Astrophysics, Australian National University, Canberra, ACT~2611, Australia\\
$^{2}$Indian Institute of Technology (ISM) Dhanbad, Jharkhand-826004, India\\
}

\date{Accepted XXX. Received YYY; in original form ZZZ}

\pubyear{2021}

\begin{document}
\label{firstpage}
\pagerange{\pageref{firstpage}--\pageref{lastpage}}
\maketitle

\begin{abstract}
We perform a series of three-dimensional, magnetohydrodynamical (MHD) simulations of star cluster formation including gravity, turbulence, magnetic fields, stellar radiative heating and outflow feedback. We observe that the inclusion of protostellar outflows (1) reduces the star formation rate by a factor of $\sim2$, (2) increases fragmentation, and (3) shifts the initial mass function (IMF) to lower masses by a factor of \mbox{$2.0\pm0.2$}, without significantly affecting the overall shape of the IMF. The form of the sink particle (protostellar objects) mass distribution obtained from our simulations matches the observational IMFs reasonably well. We also show that turbulence-based theoretical models of the IMF agree well with our simulation IMF in the high-mass and low-mass regime, but do not predict any brown dwarfs, whereas our simulations produce a considerable number of sub-stellar objects, which are produced by dynamical interactions (ejections). We find that these dynamical interactions also play a key role for the binary separation distribution and stellar kinematics in general. Our numerical model of star cluster formation also reproduces the observed mass dependence of multiplicity. Our multiplicity fraction estimates generally concur with the observational estimates for different spectral types. We further calculate the specific angular momentum of all the sink particles and find that the average value of $1.5 \times 10^{19}\, \mathrm{cm^2\, s^{-1}}$ is consistent with observational data. The specific angular momentum of our sink particles lies in the range typical of protostellar envelopes and binaries. We conclude that the IMF is controlled by a combination of gravity, turbulence, magnetic fields, radiation and outflow feedback.
\end{abstract}

\begin{keywords}
ISM: clouds -- ISM: kinematics and dynamics -- magnetohydrodynamics (MHD) -- stars: formation
\end{keywords}



\section{Introduction}
Our understanding of the star formation process has improved remarkably in past few years by virtue of advanced, high-resolution observations, rigorous theoretical works and ever-expanding numerical techniques. However, we are far from having achieved a complete picture of star formation. Numerical simulations enable scrutiny of observations and provide a framework for theoretical analysis. However, performing large-scale simulations of the collapse of molecular clouds is challenging because it involves complex, interrelated physical mechanisms like gravity, magnetic fields, and stellar radiative and mechanical feedback \citep{2018PhT....71f..38F,2019FrASS...6....7K}. Moreover, the interstellar medium (ISM) is turbulent down to the scales of molecular clouds and star-forming sub-regions \citep{2004ARA&A..42..211E}. Numerical models of star formation strive to reproduce or explain, but not limited to, the observed initial mass function (IMF), star formation rate and efficiency, multiplicity and mass ratio in stellar systems, the slow rotation rates of zero-age main sequence stars (ZAMS), and how these properties are affected by different physical mechanisms \citep[e.g.,][]{2004A&A...423....1J,2008A&A...477...25H,2009ApJ...703..131O, 2012ApJ...761..156F,2012MNRAS.419.3115B,2012ApJ...754...71K,2013ApJ...763...51F,2014MNRAS.439.3420M,2018MNRAS.476..771C,2020MNRAS.496.5072G,2020MNRAS.496.5201M,2021MNRAS.500.3594R}. Here we aim to determine the role of the combined effects of gravity, turbulence, magnetic fields, and feedback for the IMF.

The IMF represents the distribution of the mass of stars formed in a single cloud-collapse event and it is found to be surprisingly universal, not just in the solar neighbourhood, but also beyond \citep[see the reviews by][]{2010ARA&A..48..339B,2018PASA...35...39H,2020SSRv..216...70L}. \citet{1955ApJ...121..161S} proposed that the IMF can be described by a power-law of the form $dN \propto M^{-\Gamma}\, d\mathrm{log}M$, where $N$ is the number of stars, $M$ is the stellar mass and $\Gamma=1.35$. Later it was realised that the mass spectrum flattens at masses less than $1\, \mathrm{M_\odot}$ and that the distribution of low-mass stars can be represented by a log-normal function \citep{1979ApJS...41..513M}. The most commonly referenced functional forms of the IMF include a log-normal function that transforms into a Salpeter-like power-law at high masses \citep{2005ASSL..327...41C}, and a three-segment power-law \citep{2001MNRAS.322..231K}. The IMF has a peak or characteristic mass at around $0.2-0.3\, \mathrm{M_\odot}$ \citep{2003PASP..115..763C,2008ApJ...681..365E,2014prpl.conf...53O}. It is intriguing that supersonic turbulence, which embodies the chaotic nature of star formation, is also considered primarily responsible for the observed universality of the initial mass function (IMF). The role of turbulence in the star formation process is two-fold. Turbulence is supersonic on the scales of molecular clouds, and the turbulent pressure supports the cloud against a monolithic collapse \citep{1993ApJ...419L..29E,1995MNRAS.277..377P,2000ApJ...535..887K}. On the other hand, it also seeds star formation by creating local density fluctuations or over-dense regions (analogous to the observed dense cores) \citep{2004RvMP...76..125M,2007ARA&A..45..565M}. Supersonic turbulence results in a probability density function (PDF) of the gas density that is approximately log-normal \citep{1994ApJ...423..681V,1997MNRAS.288..145P,2007ApJ...665..416K,2008ApJ...688L..79F,2013MNRAS.436.1245F,2013MNRAS.430.1880H,2015MNRAS.448.3297F}. It is on this premise that recent IMF theories are formulated \citep{2002ApJ...576..870P, 2008ApJ...684..395H,2009ApJ...702.1428H,2012MNRAS.423.2037H,2013MNRAS.430.1653H}. These gravo-turbulent IMF models are built upon the observed correlation between the core mass function (CMF) and the IMF. The CMF has a form similar to the IMF, but is shifted to higher masses by a factor of 2--4 \citep{1998A&A...336..150M,1998ApJ...508L..91T,2000ApJ...545..327J,2007A&A...462L..17A,2006A&A...447..609S}. The mass gap between the CMF and the IMF is thought to be the result of the mass loss of the protostellar object due to jets and outflows, which is generally parameterized by a mass-independent core-to-star efficiency $\epsilon \sim 0.25 - 0.5$ \citep{2000ApJ...545..364M,2008ApJ...687..340M,2012ApJ...761..156F,2014ApJ...790..128F,2014ApJ...784...61O}. Jets and outflows not only remove a fraction of accreting material from the protostars, but also drive small-scale turbulence by injecting turbulent kinetic energy into the cloud \citep{2006ApJ...640L.187L,2011ApJ...740...36N}. They reduce the star formation rate significantly and aid the formation of new protostellar objects \citep{2014ApJ...790..128F}. Previous numerical studies suggest that the mechanical feedback indeed play a fundamental role in bringing out the observed mass scale of the IMF \citep{Li_2010,2012ApJ...754...71K,2018MNRAS.476..771C,2021MNRAS.502.3646G,2021MNRAS.500.3594R}. Thus, the incorporation of outflow feedback in numerical works is essential to produce conclusive results on the IMF.

The multiplicity of stars is a highly debated topic that is coupled to the IMF. Stars generally form in clusters \citep{2003ARA&A..41...57L} and observations suggest that the multiplicity fraction is an increasing function of primary mass \citep{2013ARA&A..51..269D}. Astronomers make use of the mass-dependence of multiplicity and the mass ratio in stellar systems to correct for the unresolved companions and extract the individual-star IMF from the system IMF. Core fragmentation due to the inherent rotation in dense cores is a viable mechanism for the formation of multiple systems. The angular momentum of the cores is acquired from the large-scale turbulent motions and the differential rotation of the galactic disc. The angular momentum transport is a long-standing problem in the field of astrophysics \citep{1978ppim.book.....S}. At least 5 to 6 orders of magnitude in specific angular momentum are lost between the evolution from a dense core to a typical star like the Sun \citep{1995ARA&A..33..199B,2013EAS....62...25B}. The processes proposed for solving the angular momentum problem include magnetic braking \citep{1980ApJ...237..877M}, disc formation \citep{1995ARA&A..33..505P} and removal by jets and outflows (disc winds) \citep{1982MNRAS.199..883B}. The multiplicity and angular momentum of stars are therefore key properties that can be used to probe theories of star formation.
 
The major aim of this paper is to investigate the impact of protostellar outflows on the star formation process, in particular on the IMF and the star formation rate. Our results are derived from multiple simulations to form a statistically significant sample of stars. We compare the mass distribution of stars formed in our simulations with observations and theoretical models of the IMF. We also study the multiplicity and angular momentum of the protostellar objects that form in our simulations and compare them with observational data.
 
In Section~\ref{sec:method}, we explain the simulation methodology, introduce the sub-grid models for stellar heating and outflow feedback, and define the initial configuration and simulation parameters. In Section~\ref{sec:results}, we study the effect of jets/outflows in the process of star cluster formation by comparing a model that includes the outflow feedback with a model with no protostellar outflows. For each of the two models, we investigate the column density and temperature morphology, evolution of dynamical quantities and the IMF of the stars formed in 10 cloud/cluster simulations. In Section~\ref{sec:IMF_comp}, we compare our IMF with the observational and theoretical models of the IMF. In Section~\ref{sec:Multi}, we discuss the multiplicity and the specific angular momenta of the stars. Limitations are discussed in Section~\ref{sec:discussions}. Section~\ref{sec:conclude} summarises the main results and presents our conclusions.

\section{Methods}
\label{sec:method}
\subsection{Magnetohydrodynamical equations}
The cloud-collapse is modelled numerically by solving the magnetohydrodynamical (MHD) equations including gravity on an adaptive mesh refinement (AMR) \citep{1989JCoPh..82...64B} grid using the PARAMESH library \citep{MACNEICE2000330} in the \textsc{flash4} code \citep{2000ApJS..131..273F,2008ASPC..385..145D},

\begin{equation}
\frac{\partial\rho}{\partial t}+\nabla\cdot (\rho \mathbf v) = 0,
\end{equation}
\begin{equation}
\rho\, (\frac{\partial}{\partial t} + \mathbf v \cdot \nabla )\, \mathbf v = \frac{(\mathbf B \cdot \nabla) \mathbf B}{4 \pi} - \nabla P_{\mathrm{tot}} + \rho (\mathbf g + \mathrm{\mathbf{F}_{stir}}), \label{eq:mhd2}
\end{equation}
\begin{equation} 
\frac{\partial \mathbf B} {\partial t} = \nabla \times (\mathbf v \times \mathbf B),   \hspace{4mm}\nabla \cdot \mathbf B = 0,
\end{equation}
where $\rho,\mathbf v, \mathbf B, P_{\mathrm{tot}} = P + 1/(8\pi) |\mathbf B |^2, $ and $\mathrm{\mathbf{F}_{stir}}$ denote the gas density, velocity, magnetic field, pressure (sum of thermal and magnetic) and turbulent acceleration field, respectively. Here $\mathbf{g}$ is the gravitational acceleration and is the aggregate of the self-gravity of the gas and the acceleration due to the presence of sink particles (see \S\ref{sec:sink}). We utilize a multi-grid Poisson solver for computing the self-gravity of the gas \citep{2008ApJS..176..293R}. We use the 5-wave HLL5R approximate Riemann method to solve the MHD equations. The HLL5R solver has comparable accuracy to the standard FLASH-Roe solver, but it is more efficient and has a better stability \citep{2011JCoPh.230.3331W}.

\subsection{Turbulence}
\label{sec:turb}
 The kinetic energy of freely decaying turbulence dissipates on timescales shorter than a dynamical timescale \citep{1996ApJ...466..814G,1998PhRvL..80.2754M,1998ApJ...508L..99S}. Stellar feedback, such as stellar winds and supernova explosions, inject turbulent kinetic energy and replenish the turbulence. Turbulence can also be induced by accretion and shear motions on galactic scales \citep{2009IAUS..254..289E,2017IAUS..322..123F}. We use a stochastic Ornstein-Uhlenbeck process \citep{1988CF.....16..257E,schmidt,2010A&A...512A..81F} to construct an acceleration field $\mathrm{\mathbf{F}_{stir}}$ and continuously drive turbulent motions. $\mathrm{\mathbf{F}_{stir}}$ is added as a momentum and energy source term in the MHD equations (see Eq.~\ref{eq:mhd2}). Our turbulence driving module inputs kinetic energy on the largest scales (wavenumbers $k=1\dots 3$, where $k$ is in units of $2\pi/L$ with the side length $L$ of the computational domain), which cascades down to smaller scales, naturally producing the velocity power spectrum $\sim k^{-2}$ or equivalently a velocity dispersion -- size relation of $\sigma_v \propto \ell^{1/2}$ \citep{1981MNRAS.194..809L,2002A&A...390..307O,2004ApJ...615L..45H,2011ApJ...740..120R,2013MNRAS.436.1245F,2021NatAs...5..365F}. We define the turbulence driving parameter $\zeta=0.5$ \citep{2009A&A...494..127S,2010A&A...512A..81F} to obtain a natural mixture of forcing modes. Such a value produces a ratio of compressive forcing power to the total forcing power, of about 1/3, typical for clouds in the Milky Way disc \citep{2016ApJ...832..143F}. A mixed turbulence driving represents a combination of compressive ($\nabla \times \mathbf{F_{\mathrm{stir}}} = 0,\, \zeta \sim 0)$ and solenoidal ($\nabla \cdot \mathbf{F_{\mathrm{stir}}} = 0,\, \zeta \sim 1)$ modes of driving \citep{2008ApJ...688L..79F,2010HiA....15..404F,2012MNRAS.423.2680M,2015MNRAS.451.1380N}.

\subsection{Sink particles and AMR}
\label{sec:sink}
As the central gas density of a collapsing region increases, its internal structure  becomes complex and the timescale decreases. Eventually, it becomes extremely difficult to follow the gas evolution numerically. Sink particles are sub-grid models that are used to represent the gravitationally bound, high-density regions in cloud-collapse simulations. Depending on the requirement or maximum resolution that can be achieved, sink particles are used to model the formation and accretion of individual protostellar cores or the disc + protostar systems that form later within these cores. In addition to requiring that the gas forming a sink particle be gravitationally bound, we perform a series of tests as described by \citet{2010ApJ...713..269F} before locally converting gas to sink particles to ensure that only truly bound and collapsing gas is turned into sink particles. The sink particles are introduced and centred at the computational cell that exceeds the threshold density defined by the Jeans length,
\begin{equation}
    \rho_{\mathrm{sink}} = \frac{\pi\, c_s^2}{G\, \mathrm{\lambda_J^2}} = \frac{\pi\, c_s^2}{4\,G\, r_{\mathrm{sink}}^2},
\end{equation}
where $c_s^2$ is the sound speed, $G$ is the gravitational constant, $\mathrm{\lambda_J}=[\pi c_s^2/(G\rho)]^{1/2}$ is the local Jeans length, and $r_{\mathrm{sink}}= \lambda_\mathrm{J}/2$ is the sink particle radius.

We define the size of the sink particle as $2r_{\mathrm{sink}}=5\, \Delta x$, where $\Delta x$ is the grid cell length on the highest level of refinement. This ensures that the \citet{1997ApJ...489L.179T} criterion is satisfied, avoiding artificial fragmentation. On all other AMR levels, $\lambda_\mathrm{J}$ is always resolved with at least 16 grid cell lengths to prevent underestimation of the turbulent energy and to resolve the local collapse reasonably well \citep{2011ApJ...731...62F}.

The conservation laws are exercised to update the mass, linear momentum and angular momentum of each sink particle in every accretion step. The new position of the sink particle after accretion is the centre of mass of the sink particle and the accreted material. In order to conserve total angular momentum, an intrinsic angular momentum (spin) is introduced for the sink particle, which stores the accreted angular momentum. The spin is then used to determine the rotational axis of the sink particle along which jets and outflows are launched \citep[see][]{2014ApJ...790..128F}; see details in Sec.~\ref{sec:feedback}.

All gravitational interactions of the sink particles with the gas and between each other are calculated by direct summation over all the sink particles and grid cells. A second-order leapfrog integrator is employed for advancing the sink particles in time.

\subsection{Equation of state (EOS)}
\label{sec:polytropic}
The thermodynamics of the gas in protostellar cores is determined by the competition between heating and cooling mechanisms, e.g., compressional heating, cosmic-ray heating, and cooling by dust grains \citep{1973FCPh....1....1L,1998ApJ...495..346M}. The early phase of the collapse of molecular cloud cores (birthplace of stars) is isothermal \citep{1995ApJ...443..152W,2000ApJ...531..350M,2010MNRAS.404....2G}. The cores are optically thin initially, and the gravitational energy released is readily radiated away. However, cooling becomes inefficient and the compressional heating dominates when an opaque region forms in the centre, trapping the infrared radiation from dust grains. To accurately model the thermal evolution of the gas, the equation of energy conservation has to be solved simultaneously with the RT equation. Solving the RT equation, even on the small scales of cloud cores, is difficult because it involves tracing rays that get absorbed, emitted and scattered by the constituents of the dust and gas. Further, it requires knowledge of the dust chemistry. Since we follow the collapse of molecular clouds that contain $\sim$ 20--50 cores, and considering the fact that we perform multiple simulations for better statistics, solving the energy conservation and RT equations is impractical. Therefore, to close the system of MHD equations, we use a polytropic equation of state for the gas pressure $P=P_\mathrm{EOS}$, with
\begin{equation}
    P_\mathrm{{EOS}} = c^2_s\, \rho^{\gamma}.
\end{equation}
Using the ideal gas EOS, the respective temperature is given by
\begin{equation}
    T_{\mathrm{EOS}} =  \frac{\mu\, m_{\mathrm{H}}}{k_\mathrm{B}\, \rho}\, P_\mathrm{{EOS}} = \frac{\mu\, m_{\mathrm{H}}}{k_\mathrm{B}}\, c^2_s\, \rho^{\gamma-1}\,.
\end{equation}
Here $c^2_s=(0.2\,\mathrm{km/s})^2$ is the square of the sound speed in the isothermal regime ($\gamma = 1$) for solar-metallicity, molecular gas at $10\,\mathrm{K}$, and $\mu = 2.35$ is the mean molecular weight (in units of hydrogen atom mass $m_{\mathrm{H}}$). The polytropic exponent is then set differently depending on the density of the gas, as
\begin{equation}
    \gamma =
      \begin{cases}
        1   & \text{for \hspace{7mm} $\rho \le \rho_1 \equiv 2.50 \times 10^{-16}\, \mathrm{g\, cm^{-3}}$,}\\ 
        1.1 & \text{for\, $\rho_1 < \rho \le \rho_2 \equiv 3.84 \times 10^{-13}\, \mathrm{g\, cm^{-3}}$,}\\ 1.4 & \text{for\, $\rho_2 < \rho \le \rho_3 \equiv 3.84 \times 10^{-8}\, \mathrm{g\, cm^{-3}}$,}\\
        1.1 & \text{for\, $\rho_3 < \rho \le \rho_4 \equiv 3.84 \times 10^{-3}\, \mathrm{g\, cm^{-3}}$,}\\
        5/3 & \text{for  \hspace{7mm} $\rho > \rho_4$.}
      \end{cases}
\end{equation} 
The value of the polytropic exponent $\gamma$ varies with the local density of the gas, and is based on previous detailed radiation-hydrodynamic simulations of protostar formation. It covers the phases of isothermal collapse, adiabatic heating during the formation of the first and second core and the impacts of $\mathrm{H_2}$ dissociation in the second collapse \citep{1969MNRAS.145..271L,1993ApJ...411..274Y,2000ApJ...531..350M,2009ApJ...703..131O}. However, it does not account for the increase in thermal pressure due to protostellar heating, which we discuss next.

\subsection{Stellar feedback} \label{sec:feedback}
\subsubsection{Radiative heating}
\label{sec:radfeedback}
Stellar heating feedback influences the number and mass distribution of stars formed in the collapse of molecular clouds \citep{2009MNRAS.392.1363B,2011ApJ...740...74K,2016MNRAS.458..673G,2017JPhCS.837a2007F,2020MNRAS.496.5201M,2020arXiv201003539H}. The high luminosities of young stars suppress fragmentation, allowing the stars to accrete more gas and grow in mass. Therefore, it is essential to account for the change in gas temperature due to the stellar radiative heating. But as mentioned above, solving the RT equation involving every point in space and for every timestep is computationally expensive, and therefore including it for parameter studies in large-scale simulations is often not possible. To include protostellar heating in our simulations, we use the polar stellar heating model implemented in \citet{2020MNRAS.496.5201M}. The model takes into account the existence of optically-thick accretion discs around the new-born stars and the resulting shielding of stellar radiation by dust. Based on the works of \citet{2004A&A...417..793P} and \citet{2016NewA...43...49B}, it assumes a disc density distribution around each sink particle (protostar) that is dependent on the radial distance $r$ and the angle $\theta$ subtended from the sink particle's angular momentum axis. The stellar radiant power is distributed on the grid cells surrounding the sink particle based on this dust/disc density distribution.

Dust particles absorb the radiation from the central star with the rate of energy absorption given by
\begin{equation}
    Q (r, \theta) = \chi\, \frac{L_{\star}}{4\pi r^2}\, \exp\left(-\tau (r,\theta)\right), \label{eq:Qheat}
\end{equation}
where $\chi$ is the absorption coefficient. The star's luminosity ($L_{\star}$), which includes both the accretion and intrinsic luminosities, is calculated by employing the protostellar evolution model by \citet{2009ApJ...703..131O}. The total optical depth ($\tau$) in any direction given by $\theta$ is
\begin{equation}
    \tau = \int \kappa\, \rho(r,\theta)\, \mathrm{d}r, 
\end{equation}
where $\kappa$ is the grey opacity and $\rho(r,\theta)$ is the assumed dust/disc density distribution \citep[see][for a description of the analytical model of the disc density distribution used here]{2020MNRAS.496.5201M}. The radiation is attenuated in the directions of the disc because of the absorption by dust grains, and therefore the main heating is confined to the polar directions.

At the equilibrium temperature, the amount of energy emitted by the dust grains will be equal to the amount they absorb, i.e., $Q$. Therefore we can write
\begin{equation}
    \frac{\sigma_\mathrm{{SB}}}{\pi}\, \chi\, T_{\mathrm{heat}}^4 = \frac{Q}{4\pi},
\end{equation}
where $\sigma_\mathrm{{SB}}$ is the Stefan-Boltzmann constant and $T_{\mathrm{heat}}$ is the temperature due to stellar radiative heating. Here we have neglected the reprocessed radiation field.

To account for the change in thermal pressure due to stellar radiative feedback, we add the space-dependent pressure term obtained from the polar stellar heating module to the pressure computed from the polytropic equation of state \citep[see][]{2016MNRAS.458..673G,2018AAS...23111403G,2017JPhCS.837a2007F}. Thus, the final gas pressure is
 \begin{align}
    P &= \left[P^4_{\mathrm{EOS}} + P^4_{\mathrm{heat}}\right]^{1/4} \nonumber \\ 
      &= \left[P^4_{\mathrm{EOS}} + {\left(\frac{k_B\, \rho}{\mu\, m_\mathrm{H}}\right)}^4\, T^4_{\mathrm{heat}}\right]^{1/4}, \label{eq:Pheat}
\end{align}
which is imposed in the MHD momentum equation, Eq.~(\ref{eq:mhd2}).

\subsubsection{Jets/Outflows}
\label{sec:outflowfeedback}
The bipolar mechanical feedback from stars consists of two components \textemdash\, the highly collimated fast stream of gas, called jets, that drill through the cloud, and the wide-angle low-speed molecular outflows \citep{2014prpl.conf..451F}. We employ the subgrid-scale (SGS) outflow model of \citet{2014ApJ...790..128F} for launching this jet/outflow combination in our simulations. It captures the fast jet component and includes angular momentum transfer.

The SGS module imparts momentum uniformly to the grid cells within a confined volume defined by two spherical sections (cones) around the sink particle. The cones open towards the opposite poles of the sink particle and are characterised by an opening angle $\theta_{\mathrm{out}}=30^{\circ}$ measured from the angular momentum axis. We set the radial extent equal to $r_{\mathrm{out}}=16\Delta x$ measured from the position of the sink particle (tip of the cone), where $\Delta x$ is the cell size on the highest level of refinement, as tested and recommended in \citet{2014ApJ...790..128F}. The model uses radial and angular smoothing functions to prevent numerical artifacts at the interfaces. The momentum injected into each of the cones is 
\begin{equation}
    \mathbf{P_{\mathrm{out}}}=\pm (1/2)\, M_{\mathrm{out}}\, \mathbf{V_{\mathrm{out}}},
\end{equation}
where $M_{\mathrm{out}}$ is the ejected mass, which is equal to the fraction $f_{\mathrm{m}}$ of the mass accreted by the sink particle in a timestep $\Delta t$, i.e., $M_{\mathrm{out}} = f_{\mathrm{m}}\, \dot M_{\mathrm{acc}}\, \Delta t$. We use $f_{\mathrm{m}}=0.3$ \citep{2014ApJ...790..128F}, which is consistent with observations \citep{1995AJ....109.1846H,2007A&A...468L..29C,2011ApJ...737L..26B}, theoretical models of the outflow mechanism \citep{1982MNRAS.199..883B,1988ApJ...328L..19S,2007prpl.conf..277P} and the values derived from previous numerical simulations \citep{2008A&A...477....9H,2012MNRAS.422..347S,2013ApJ...774...12F}

$\mathbf{V_{\mathrm{out}}}$ is set to the Kepler speed near the surface of the protostar, such that
\begin{equation}
    |\mathbf{V_{\mathrm{out}}}| = 100\, \mathrm{km\, s^{-1}} \left(\frac{M_{\mathrm{sink}}}{0.5\, \mathrm{M_\odot}}\right)^{1/2},
\end{equation}
where $M_{\mathrm{sink}}$ is the mass of the sink particle and $100\, \mathrm{km\, s^{-1}}$ is the typical jet speed (and Kepler speed) for a protostar of mass $M = 0.5\, \mathrm{M_\odot}$ at a radius of $R = 10\, \mathrm{R_\odot}$. $\mathbf{V_{\mathrm{out}}}$ consists of a slow component with a speed of $0.25\, |\mathbf{V_{\mathrm{out}}}|$ and a fast component with speed of $0.75\, |\mathbf{V_{\mathrm{out}}}|$. The high-speed component contributes to the momentum injection in the cones only within an opening angle of $5^{\circ}$. By using such a velocity profile, the model distinguishes the faster jet and the slower molecular outflow components. A fraction $f_\mathrm{a}$ of the accreted angular momentum is removed from the sink particle and re-introduced to the two outflow/jet components. We use the default value of $f_\mathrm{a}=0.9$ in the SGS model, which is based on the observations in \citet{2002ApJ...576..222B} and numerical works, e.g., \citet{2006ApJ...641..949B} and \citet{2008A&A...477....9H}. 

The inserted momentum into the two cones is effectively carried away to larger scales by the MHD code. The SGS outflow model reproduces converged large-scale outflow properties for the mass, linear momentum, angular momentum and the outflow speed, almost independent of the resolution. We refer the reader to \citet{2014ApJ...790..128F} and references therein for a detailed description of the SGS model and justification of the parameter choices. 

\subsection{Initial conditions and simulation parameters}
\label{sec:parameters}

The computational domain of the simulations is a three-dimensional triple-periodic box with side length $L=2\, \mathrm{pc}$. The maximum refinement level provides a maximum effective grid resolution of $N_{\mathrm{eff,\, res}}^3=4096^3$ cells or a minimum cell size of $\Delta x_{\mathrm{cell}}=100\,\mathrm{AU}$. The initial gas density is uniform with $\rho_{\circ} = 6.56 \times 10^{-21}\, \mathrm{g\, cm^{-3}}$, which gives a total cloud mass of $M_{\mathrm{cl}}=775\, \mathrm{M_{\odot}}$  and a mean free-fall time of $t_{\mathrm{ff}}= 0.82\,$Myr. The turbulence driving module stirs the gas in the box initially in the absence of self-gravity. The induced turbulence creates cloud-typical structures and density contrasts in the form of filaments and clumps. The high-density regions within these structures are potential sites of star formation \citep{2011A&A...529L...6A,2013A&A...551C...1S,2014prpl.conf...27A}. The amplitude of the turbulence driving is set by the velocity dispersion $\sigma_v=1.0\, \mathrm{km\, s^{-1}}$ and the initially isothermal sound speed $c_s=0.2\, \mathrm{km\, s^{-1}}$, which gives a steady-state sonic Mach number of $\mathcal{M}=\sigma_v/c_s=5.0$. The magnetic field, which is uniform initially with $B= 10^{-5}\, \mathrm{G}$ along the z-axis of the computational box, is later modified due to the compression, tangling and folding by the turbulence, approximating the morphology of magnetic fields in real molecular clouds \citep{2016JPlPh..82f5301F}. The initial virial parameter $\alpha_\mathrm{{vir}}=2E_\mathrm{{kin}}/E_\mathrm{{grav}}=0.5$ is in the range of observed values \citep{1992A&A...257..715F,2013ApJ...779..185K,2015ApJ...809..154H}. After two turbulent crossing times, $2 t_\mathrm{turb}=L/(\mathcal{M}c_s) = 2\,\mathrm{Myr}$, a fully-developed turbulent state is reached , which is when we turn on self-gravity and allow for sink particle creation, which is the typical procedure for this type of cluster-formation experiments. We study the temporal evolution of different dynamical quantities and statistical properties of the forming star clusters from this point in time, which we define as $t = 0$, i.e., when self-gravity is activated. This method is similar to that used in earlier works \citep[e.g.,][]{2012ApJ...761..156F,2012ApJ...754...71K,2016ApJ...822...11P,2018MNRAS.480..182G}.

\begin{table*}
	\caption{Key simulation results.}
	\label{tab:sims}
	\begin{tabular}{lcccccccc} 
	    \hline
		\hline
		 Model & $t_{5\%} [t_{\mathrm{ff}}]$ & $\overline{\mathrm{SFR}}_{\mathrm{ff}} [\%]$ & $N_{\mathrm{Total\, sinks}}$ & $\overline{M}_{\mathrm{median}}\, [\mathrm{M_{\odot}}] $ & $\overline{M}_{\mathrm{avg}}\, [\mathrm{M_{\odot}}]$ & $\Gamma$ & SSF & $j_\mathrm{mean}\, [\mathrm{cm^2\, s^{-1}}]$\\
        (1) & (2) & (3) & (4) & (5) & (6) & (7) & (8) & (9)\\
		\hline
		\hline 
		1. NOWIND  & $0.68\pm0.15$ & $15\pm3$ & 212 & $1.0\pm0.3$ & $1.9\pm0.3$ & $1.3\pm 0.2$ & $0.65$ & $1.2 \times 10^{20}$\\
		2. OUTFLOW & $0.89\pm0.20$ & $7\pm2$ & 449 & $0.5\pm0.1$ & $0.9\pm0.1$ & $1.5\pm 0.3$ & $0.66$ & $1.5 \times 10^{19}$\\
		
		\hline
	\end{tabular}
	\\
    \raggedright\textbf{Notes.} Ten simulations with different turbulence realisations (T1--T10) are run for both the NOWIND and OUTFLOW models. In the table, $t_{5\%}$ is the average time taken (in units of the free-fall time) by the simulations to reach SFE = 5\% and is measured from the moment self-gravity is turned on. The value of $\overline{\mathrm{SFR}}_{\mathrm{ff}}$ quoted in the table is time averaged, while all the other quantities are calculated at SFE = 5\%. The resolution level, cloud properties and turbulence setup are the same in both models and the only difference is that protostellar outflows are absent in the NOWIND simulations. Main simulation parameters: computational box size: $L=2\, \mathrm{pc}$, maximum effective grid resolution: $N_{\mathrm{eff,\, res}}^3=4096^3$ cells, minimum cell size: $\Delta x_{\mathrm{cell}}=100\,\mathrm{AU}$, sink particle threshold density: $\rho_{\mathrm{sink}}=3.8\times10^{-16}\, \mathrm{g\ cm^{-3}}$, uniform initial gas density: $\rho_{\circ} = 6.56 \times 10^{-21}\, \mathrm{g\, cm^{-3}}$, total cloud mass: $M_{\mathrm{cl}}=775\, \mathrm{M_{\odot}}$, uniform initial magnetic field: $B=10^{-5}\, \mathrm{G}$ (along the z-axis), turbulence driving parameter: $\zeta=0.5$, velocity dispersion on the driving scale of the turbulence: $\sigma_v=1.0\, \mathrm{km\, s^{-1}}$.
\end{table*}

\section{Results}
\label{sec:results}
We run 10 simulations of star cluster formation with different turbulence realisations (T1--T10), incorporating gravity, turbulence, magnetic fields, stellar heating and outflow feedback (Model OUTFLOW).  We evolve the simulations till 5\% of the total mass of the cloud has formed stars, i.e., the star formation efficiency SFE $=5\%$. The objective of performing multiple simulations is to produce a sufficient number of stars to obtain statistically conclusive results. Since we include a set of physical mechanisms (gravity, turbulence, magnetic fields, and both mechanical and radiation feedback) governing the star formation process, we are in a good position to compare the statistical properties like the IMF and multiplicity with observations and theory. We begin by examining how the evolution of the cloud is influenced by the outflow feedback.

\begin{figure*}
    \centering
    \includegraphics[width=\textwidth]{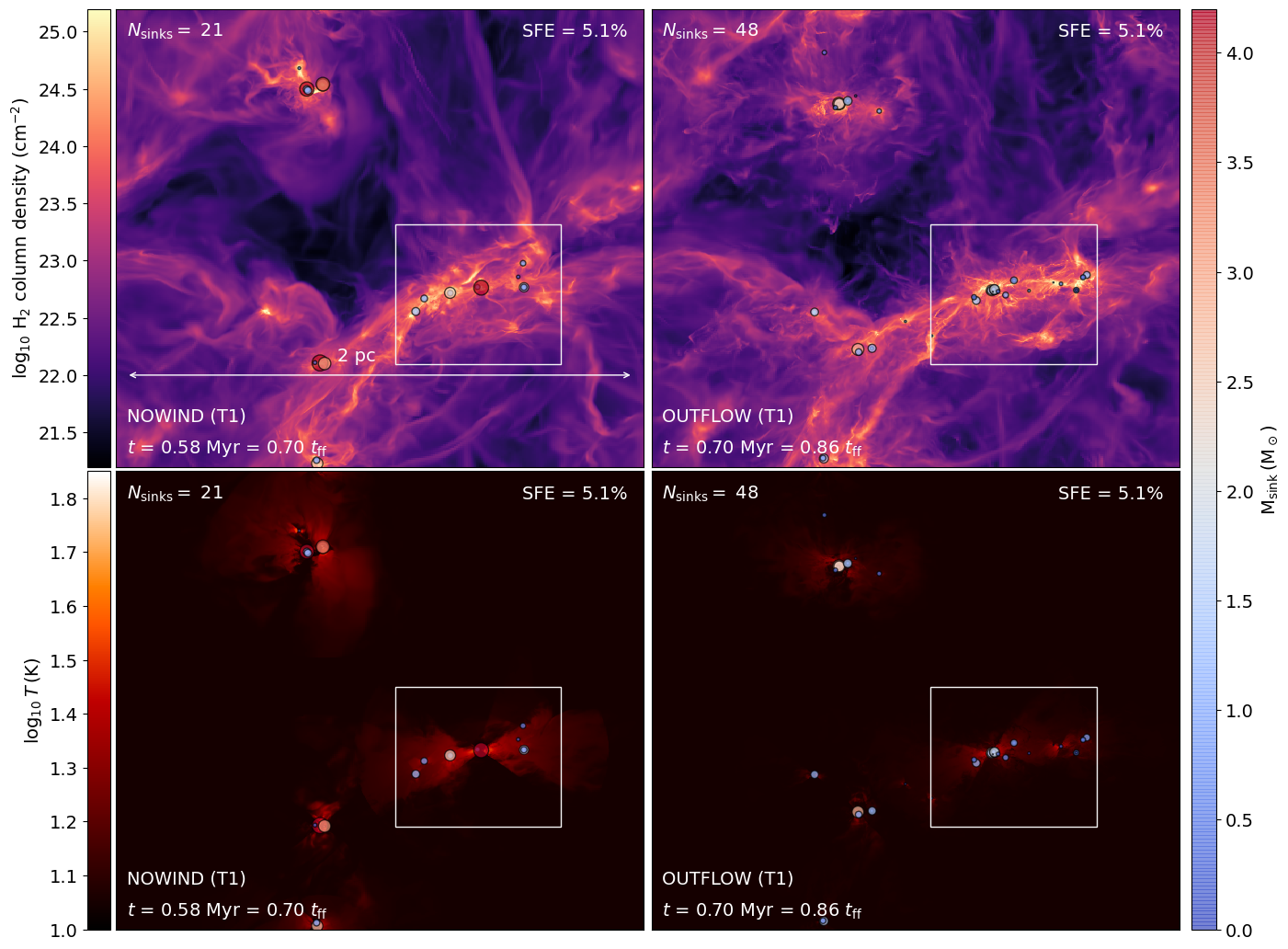}
    \caption{The first column presents the mass-weighted\protect \footnotemark[1] projection maps of the gas number density (top) and temperature (bottom) of the NOWIND model and the second column shows the same for the OUTFLOW model at SFE = 5\%. The circular markers in each panel represent the position of the sink particles formed in the simulations. The colour and size of the markers are scaled by the mass of the sink particles (see right-hand colour scale). The size of the markers should not be confused with the numerical size of the sink particles, which is constant with a radius of $r_{\mathrm{sink}}= 250\, \mathrm{AU}$. The OUTFLOW model produces 48~sinks compared with 21~sinks in the NOWIND model. The SFR during the main star formation stages is reduced by a factor $\sim2$ in the OUTFLOW model compared to the NOWIND case (see Tab.~\ref{tab:sims}), and there is less heating in the OUTFLOW model (compare the bottom panels).}
    \label{fig:t1_samesfe}
\end{figure*}

We investigate the impact of jets/outflows on the number and distribution of the stars formed and the star formation rate (SFR) by comparing the simulations of the OUTFLOW model with another set of 10 simulations with no outflow feedback (Model NOWIND). The simulation setup and turbulence seeds used in the NOWIND simulations are the same as in the OUTFLOW simulations. Fig.~\ref{fig:t1_samesfe} shows the mass-weighted column density (integral of the number density weighted by mass along the line-of-sight) of both the models for one particular turbulence realisation (simulation~T1). The NOWIND and OUTFLOW models form 21 and 48 sink particles, respectively. The NOWIND simulation reaches an SFE $=5\%$ in 0.58~Myr, while the OUTFLOW simulation takes 0.70~Myr to reach the same SFE. It is evident from the temperature maps (bottom panel in Fig.~\ref{fig:t1_samesfe}) that the gas temperature around the sink particles in the NOWIND model is higher than that in the OUTFLOW simulation. 

\begin{figure*}
    \centering
    \includegraphics[width=\textwidth]{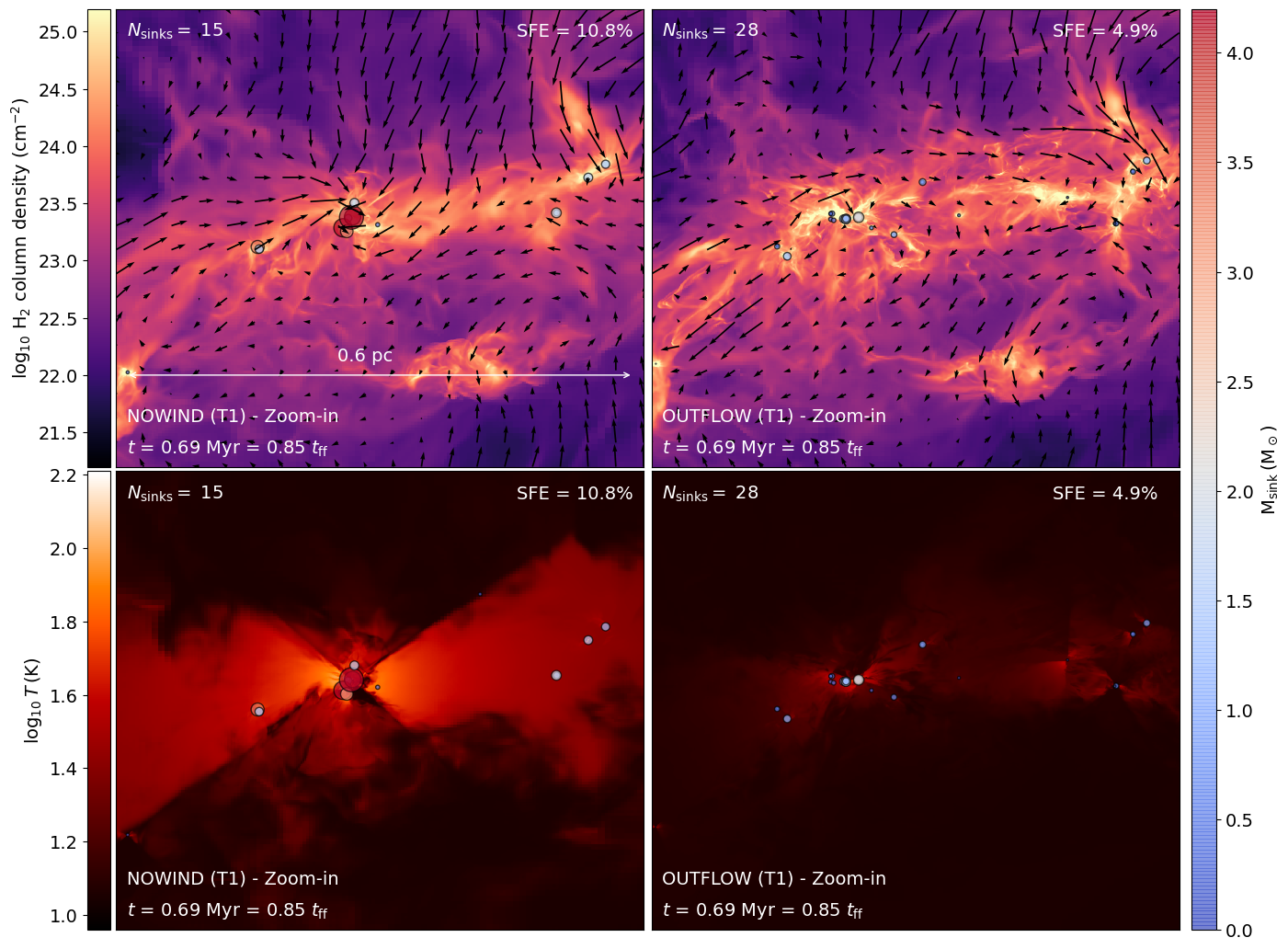}
    \caption{Zoom-in images of the region within the squares in Fig.~\ref{fig:t1_samesfe} showing the contrast in the morphology and temperature structure between the NOWIND (left) and OUTFLOW (right) models at the same simulation time. The arrows represent the velocity field lines of the gas.}
    \label{fig:t1_sametime}
\end{figure*}

\footnotetext[1]{We define the mass-weighted projection of the gas number density as $\int \rho^2\, dz\, / \int \rho dz$ and the mass-weighted projection of the temperature as $\int \rho T dz\, / \int \rho dz$, where the projection is taken along the $z$-direction. All figures in this paper showing density and temperature maps are mass-weighted. The aim of the mass-weighting is to enable better visualisation of the morphological features, i.e., to bring out the densest structures. A version of Fig.~\ref{fig:t1_sametime} without mass-weighting is shown in Fig.~\ref{fig:t1_sametime_nomw}.}

The OUTFLOW model generates a higher number of stars as compared to the NOWIND model at the same SFE of 5\% because (1) the molecular cloud in the OUTFLOW model has evolved further in time, and therefore some stars that form independent of the presence of the outflow feedback have not yet emerged in the NOWIND case, (2) the outflows from protostellar objects can inject enough momentum into the cloud to drive small-scale turbulence and perturb the accretion flow, allowing the formation of more stars \citep{2010ApJ...709...27W,2014ApJ...790..128F}, and (3) they indirectly lower the efficiency of stellar heating. The ejection of the accreted material from the star+disc system and the sweeping away of a part of the surrounding envelope by disc and/or stellar winds reduce the mass of the stars and therefore their luminosity \citep{2012ApJ...747...22H,2012ApJ...754...71K}. Thus, the ability of the stellar heating feedback to suppress fragmentation is significantly reduced when jet/outflow feedback is included.

The influence of the outflows in promoting star formation by redirecting accretion flows and in reducing the stellar heating efficiency can be inferred from Fig.~\ref{fig:t1_sametime}. The figure shows the gas density structure and the temperature distribution of the region within the marked squares in Fig.~\ref{fig:t1_samesfe} for each of the models at the same simulation time. At the time the OUTFLOW model has reached SFE = 5\%, the NOWIND model has already reached SFE = 11\%. Even in this small region of size $0.6\, \mathrm{pc}$, the OUTFLOW model has almost double the number of stars. Since the considered region of both the models are now viewed at the same simulation time $t=0.69\, \mathrm{Myr}$, we can say with confidence that the increase in the number of sinks with the inclusion of outflows is not just because of the slow star formation rate, i.e., the OUTFLOW model being more evolved than the NOWIND model when considered at the same SFE (see also panel (c) in Fig.~\ref{fig:dynamicQ}). The stars in the OUTFLOW model are relatively lower in mass, and also the main filament in which most of the stars form in this region breaks into several sub-fragments due to the action of the jets/outflows. By contrast, the NOWIND model preserves the main filament structure with significantly less fragmentation, and the matter within the main filament is accreted by the existing stars. Clearly, the stellar heating is more efficient in the NOWIND model. The gas temperature in the immediate regions around the high-mass stars in the NOWIND simulation reach a few hundred Kelvin, which leads to suppression of fragmentation.

\begin{figure*}
    \centering
    \includegraphics[width=\textwidth]{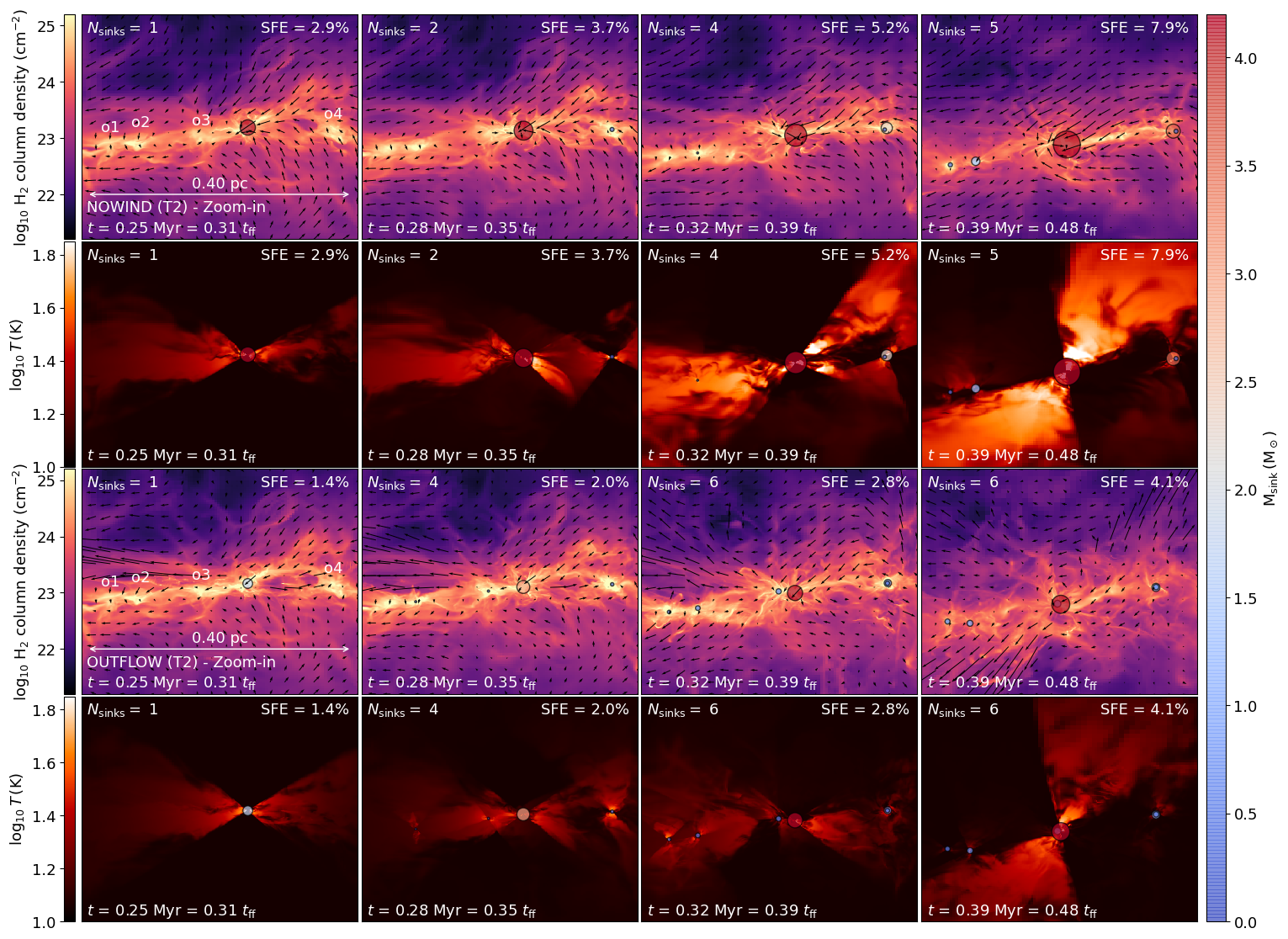}
    \caption{Zoom-in time snapshots of a filament in the T2 simulation of the NOWIND (rows 1 and 2) and the OUTFLOW (rows 3 and 4) models showing the evolution of over-densities within the filament and comparing the evolution of the formation (or absence thereof) of stars in locations o1, o2, o3, and o4, as labelled in the first panel of each simulation set.}
    \label{fig:t2_compare}
\end{figure*}

\begin{figure*}
    \centering
    \includegraphics[width=\textwidth]{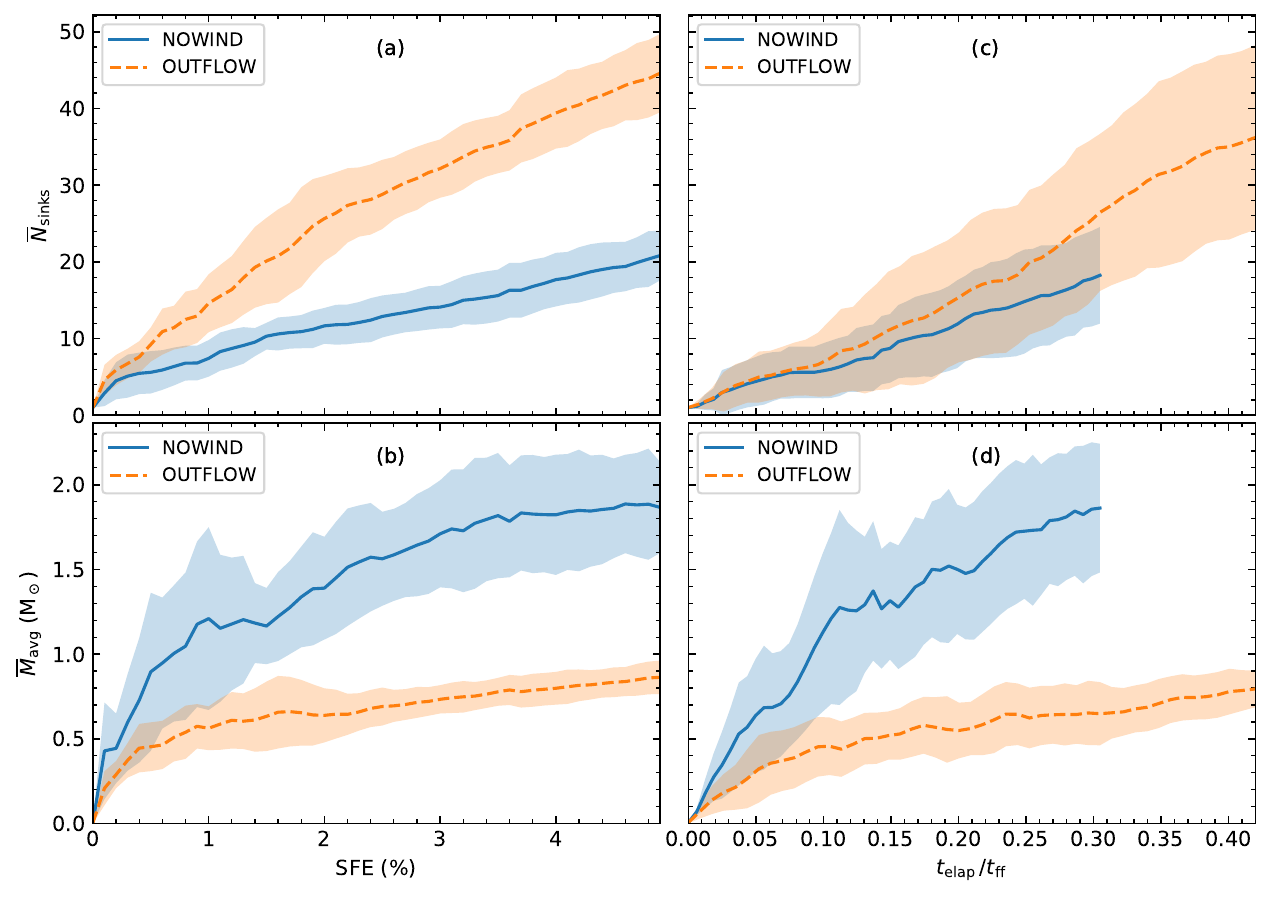}
    \caption{The left panels show (a) the average number of sink particles formed and (b) the average stellar mass as a function of the star formation efficiency (SFE in \%). The right panels (c) and (d) indicate the average number of sink particles formed and the average stellar mass, respectively, as a function of time. All quantities shown here for both the models correspond to the average values obtained from 10~simulations (T0--T10), and the coloured bands represent the standard deviation over the set of these 10~simulations. Here $t_{\mathrm{elap}}/t_{\mathrm{ff}}$ is the elapsed time from the formation of the first sink particle in units of the free-fall time and is distinguished from the time $t$ in the above column density maps, which is the time measured from the moment self-gravity was turned on.} 
    \label{fig:dynamicQ}
\end{figure*}

Fig.~\ref{fig:t2_compare} presents the time evolution of a 0.4~pc sub-region in the cloud of NOWIND and OUTFLOW simulations with turbulence realisation T2. At $t = 0.25\, \mathrm{Myr} = 0.31\, t_\mathrm{ff}$ (first panel), there are four over-dense regions close to the already formed sink particle, marked as o1, o2, o3 and o4. The sink particle in the OUTFLOW case is lower in mass than the one in the NOWIND case because of the mass loss through jet/outflows. The outflow axis of the sink is parallel to the filament, which can be inferred from the velocity field lines and the orientation of the heating zones around the sink particle. The radiative flux is primarily in the directions of the disc/outflow axis, because of the extinction by dust particles in the disc. It can be observed from the snapshots that the sink particles that formed in o2 and o1 in the OUTFLOW model at $t = 0.28\, \mathrm{Myr}$ and $t = 0.32\, \mathrm{Myr}$, respectively, only form much later in the NOWIND case. This suggests that protostellar outflows can trigger local star formation in nearby cores. Core o4 harbours a binary in both models, for which both components of the binary formed around the same time. In the NOWIND simulation, o3 does not form a star and all the gas is completely accreted by the high-mass sink particle. The higher gas temperature in o3 due to the higher luminosity of the massive sink particle in the NOWIND model has likely prevented the collapse of o3. However, in the OUTFLOW case, a sink particle does form in o3, which later goes into a close binary orbit with the massive sink particle. 

\begin{figure}
    \centering
    \includegraphics[width=\columnwidth]{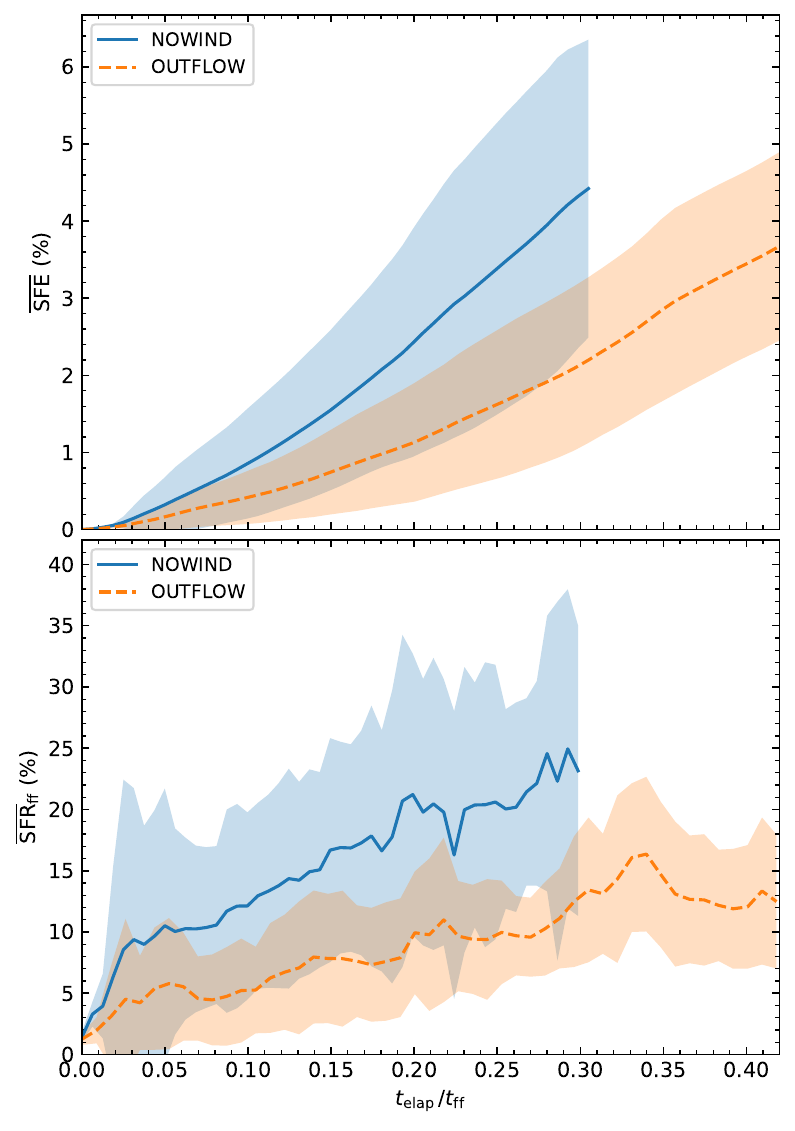}
    \caption{The top and bottom panels show the time evolution of the average SFE and average star formation rate per free-fall time ($\mathrm{SFR_{ff}}$ in \%), respectively.} 
    \label{fig:SFRplot}
\end{figure}

We plot the evolution of the average number of sink particles and the average stellar mass with SFE and time in Fig.~\ref{fig:dynamicQ}. We find that, at the same SFE, the ratio of the number of sinks formed in the OUTFLOW model to that in the NOWIND model is $2.1\pm0.1$ averaged over the SFE range 1--5\%. \citet{Li_2010} also find that their simulation including outflow feedback has twice as many stars as their simulation with no outflows when compared at the same SFE. When compared at the same time, the number of sink particles in the OUTFLOW model is higher by a factor of $1.3\pm0.1$ in the range $0.1 < t_{\mathrm{elap}}/t_{\mathrm{ff}} < 0.3$. The momentum injection by the outflows and the reduced heating effect are responsible for the difference in the number of sink particles between the models when compared at the same time. \citet{2014ApJ...790..128F} detect that the inclusion of outflow feedback in simulations increases the number of stars by a factor of $\sim 1.5$ compared to simulations with no outflows at the same time. The simulation model in \citet{2014ApJ...790..128F} does not include the stellar radiative heating and therefore the momentum injection by the outflows is solely responsible for the increase in the number of stars in their study. This, along with the fact that they also observe an increase by a similar factor as in our case, suggests that, although the reduced heating effect is important, the momentum injection by outflows plays a more significant role in increasing fragmentation. The average sink particle mass ($\overline{M}_{\mathrm{avg}}$) in the NOWIND case increases with the evolution of the cloud, while it is relatively constant with a value of $0.7\pm0.1\, \mathrm{M_\odot}$ in the OUTFLOW case. Here the measurement is made by taking the average of $\overline{M}_{\mathrm{avg}}$ over the SFE range 1--5\% with the error bars corresponding to the standard deviation over this SFE range. The overbar in $\overline{M}_{\mathrm{avg}}$ denotes the average over the 10~simulations. At the same SFE, $\overline{M}_{\mathrm{avg}}$ in the OUTFLOW simulations is lower than in the NOWIND simulations by a factor of $2.2\pm0.2$, reflecting the fact that the number of stars increased by a similar factor at the same SFE. When compared at the same time, $\overline{M}_{\mathrm{avg}}$ in the OUTFLOW simulations is lower by a factor of $2.6\pm0.2$. Fig.~\ref{fig:SFRplot} presents the average SFE and average star formation rate per free-fall time ($\mathrm{SFR_{ff}}$ in \%) \citep{2005ApJ...630..250K,2012ApJ...761..156F} as functions of time. We observe that the star formation rate per free-fall time averaged over 10 simulations, $\overline{\mathrm{SFR}}_{\mathrm{ff}}$~(\%), increases with time, but the progression is slower in the OUTFLOW model, because of self-regulation by the outflows. $\overline{\mathrm{SFR}}_{\mathrm{ff}}$ in the NOWIND and OUTFLOW simulations are $15\pm3\%$ and $7\pm2\%$, respectively, in the range 0.05 < $t_{\mathrm{elap}}/t_{\mathrm{ff}}$ < 0.25 $t_{\mathrm{ff}}$ (see Tab.~\ref{tab:sims}). Therefore, the stellar outflows reduce the SFR by a factor of $\sim2$. Our results are consistent with \citet{2014ApJ...790..128F} who also observed a reduction in the SFR by the same factor with the inclusion of outflow feedback. As mentioned above, one major difference between our model and that of \citet{2014ApJ...790..128F} is that stellar heating was missing in the latter. However, the SFR is relatively insensitive to radiative feedback \citep{2020MNRAS.496.5201M}. Although our value of $\mathrm{SFR}_{\mathrm{ff}}$ (OUTFLOW case) is higher than the average value in the Milky Way ($\sim 1$--$2\%$ per free-fall time)  \citep{2007ApJ...654..304K,2019MNRAS.488.1407K}, it is still within the dispersion of $\mathrm{SFR}_{\mathrm{ff}}$ obtained in observational surveys and from theoretical predictions \citep[e.g.,][]{2005ApJ...630..250K,2009ApJS..181..321E,2011ApJ...743L..29H,2011ApJ...729..133M,2012ApJ...761..156F,2016ApJ...831...73V,2016ApJ...833..229L}.

Finally, we mention that the properties of the turbulence can significantly affect the SFR \citep{2012ApJ...761..156F} and the IMF \citep{2010A&A...516A..25S,2021MNRAS.503.1138N}. For example, the effect of the turbulence driving and Mach number on the SFR and IMF with the physics included here will be the main focus of a follow-up study.

\subsection{Sink particle mass distributions}
\label{sec:SMD}
\begin{figure}
    \centering
    \includegraphics[width=\columnwidth]{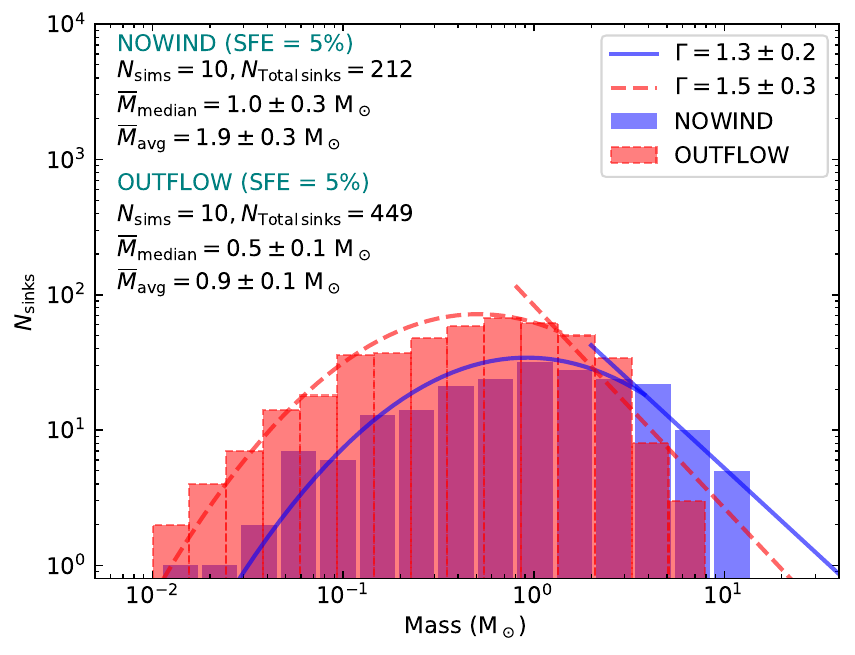}
    \caption{Comparison between the sink mass distribution of the NOWIND (no outflow feedback) and OUTFLOW models at SFE = 5\%. The data compilation in each model is acquired from 10~simulations. The histogram with solid edges and the solid curves correspond to the NOWIND model, while the histogram with dashed edges and the dashed curves correspond to the OUTFLOW model. The standard deviation of the lognormal curves are equal to that of the \citet{2005ASSL..327...41C} IMF ($\sigma=0.55$), but the peaks are located at around the median (mean) mass of the respective SMDs, i.e., at $1.0\,(1.9)\, \mathrm{M_\odot}$ for NOWIND and $0.5\,(0.9)\, \mathrm{M_\odot}$ for OUTFLOW. $\overline{M}_{\mathrm{median}}$ and $\overline{M}_{\mathrm{avg}}$ are the median and average sink masses averaged over the 10~simulations at SFE = 5\%, where the error bars represent the standard deviation (see Tab.~\ref{tab:sims}).} 
    \label{fig:imf_comp}
\end{figure}

The mass distribution of stars formed in the 10~simulations of each model is presented in Fig.~\ref{fig:imf_comp}. With the introduction of the outflow feedback, the sink mass distribution (SMD) shifts to lower masses, with the median and average masses reduced by factors of 2.0 and 2.1, respectively. However, putting aside the shift in mass, the OUTFLOW SMD has almost the same shape as that of the NOWIND SMD. This means that outflows do not seem to significantly change the basic shape (or functional form) of the IMF, and thus, the same basic mechanism(s) that set the high-mass slope of the IMF and a turnover at intermediate masses seem to be at play regardless of whether jet/outflow feedback is included or not. A log-normal function with a standard deviation for the Chabrier IMF fits both models well at the low-mass end, with a peak at $1.0\,\mathrm{M}_\odot$ for the NOWIND distribution and $0.5\,\mathrm{M}_\odot$ for the OUTFLOW SMD. While the power-law fit to the high-mass end of the OUTFLOW SMD ($\Gamma=1.5\pm 0.3$) is slightly steeper than the fit to the NOWIND SMD ($\Gamma=1.3\pm 0.2$), they are statistically identical within the 1-sigma uncertainties.

The parallelism between the two distributions resembles the observed correlation between the observed core mass function (CMF) and the stellar IMF. The existence of a correspondence between the CMF and the IMF implies that a star's mass is decided at the core level. The recognition of dense cores as the direct progenitors of stellar objects allows us to explain why the overall form of the distribution did not change with the inclusion of mechanical feedback. Supersonic turbulence creates density enhancements of varying sizes, but not all of these over-densities form stars. Only the ones that exceed the threshold mass for collapse will form stars. Outflows do not create new density enhancements; instead, they increase the chance that a star forms in an over-density generated by gravo-turbulent fragmentation. This can be seen in Fig.~\ref{fig:t2_compare}, where the extra sink particle forms in an over-density that pre-existed (see Fig.~\ref{fig:t1_compare} of Appendix A for another example). It should then naturally follow that the mass distribution of extra sinks in the OUTFLOW SMD also reflects the distribution of over-density masses, i.e., the typical mass of the extra sinks should conform to the typical over-density mass with an efficiency factor decided by the mass loss via winds and entrainment of the envelope material by jets. 

Observational surveys of dense cores find high-mass CMF slopes of $\Gamma$ between 1.0 and 1.6 \citep{1998A&A...336..150M,1998ApJ...508L..91T,2000ApJ...545..327J,2007MNRAS.374.1413N,2007A&A...462L..17A}, similar to the Salpeter slope of the IMF, but the range of CMF slopes does not impose strong constraints on the similarity of the CMF and the IMF. Nevertheless, we find that the introduction of the outflow feedback shifts the IMF to lower masses, but virtually sustaining its overall shape. Within the limit of these arguments, the present set of simulations suggests that jet/outflow feedback can be responsible for the observed correlation and shift between the CMF and the IMF \citep{2007A&A...462L..17A,2007MNRAS.379...57C,2008MNRAS.391.1091S}.

\section{Comparison with observational data and theoretical IMF models}
\label{sec:IMF_comp}
\subsection{Comparisons with observed IMFs}

\begin{figure}
    \centering
    \includegraphics[width=\columnwidth]{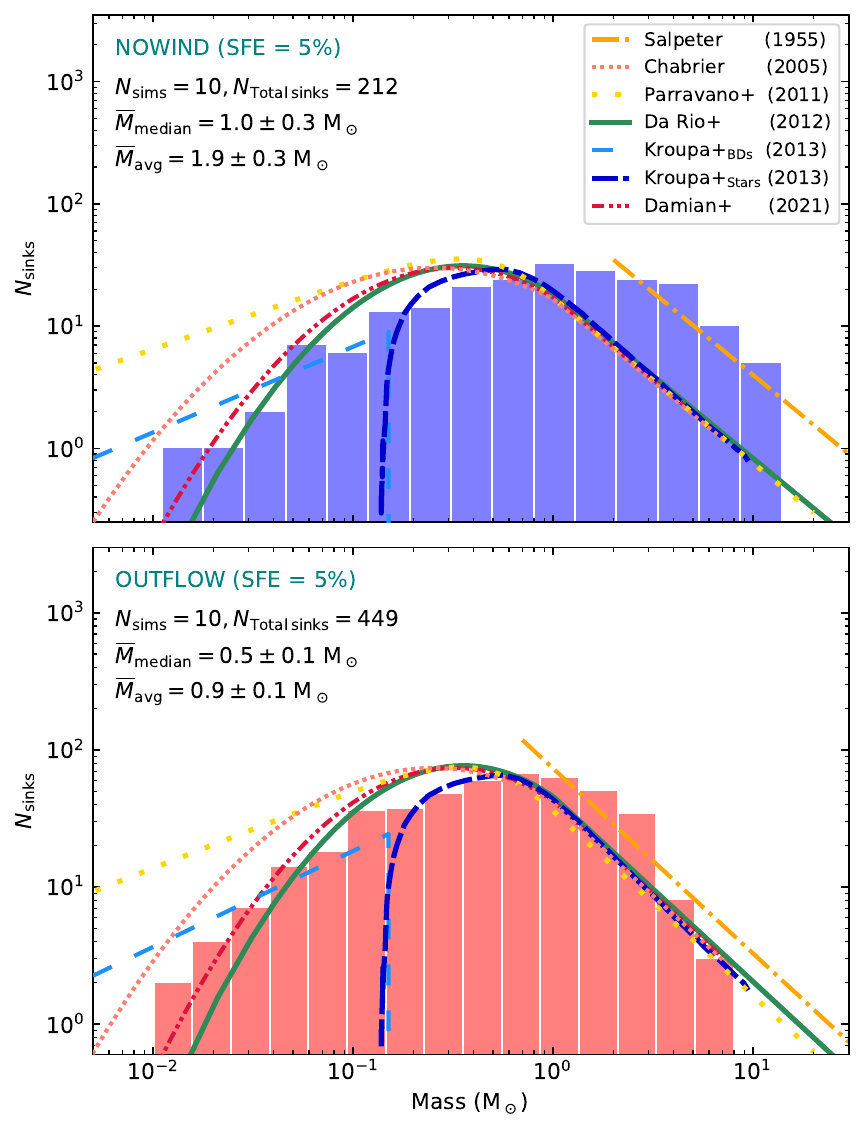}
    \caption{Top panel: Comparison between the sink mass distribution of the NOWIND model at SFE = 5\% with various observational IMFs. Bottom panel: Same as the top panel, but for the OUTFLOW data. The plotted curves are the system IMF models, based on observations, by \citet{1955ApJ...121..161S} (dash-dotted), \citet{2005ASSL..327...41C} (short-dotted), \citet{2011ApJ...726...27P} (long-dotted), \citet{2012ApJ...748...14D} (solid), \citet{2013pss5.book..115K} for brown dwarfs (long-dashed) and stars (short-dashed), and \citet{2021MNRAS.504.2557D} (dash-dot-dotted).}  
    \label{fig:imf_obs}
\end{figure}

Fig.~\ref{fig:imf_obs} compares the distribution of the sink masses formed in the 10 simulations of each model with various fits to the observed IMF studied in the literature since \citet{1955ApJ...121..161S} (dash-dotted line). We compare the SMDs with the system IMFs rather than the individual-star IMFs, because we do not resolve all the low-order multiple systems. The \citet{2005ASSL..327...41C} system IMF is represented by a short-dotted curve. The \citet{2011ApJ...726...27P} IMF (long-dotted line) is an analytical model defined by multiple parameters based on observational constraints, e.g., the slope of the high-mass end of the IMF and the ratio of the number of brown dwarfs (BDs) to the number of stars \citep[see also][]{2000ApJ...534..870P}. This function predicts a higher fraction of BDs below $0.03\, \mathrm{M_\odot}$ than the \citet{2005ASSL..327...41C} IMF. \citet{2012ApJ...748...14D} (solid line) measured the IMF in the Orion Nebula Cluster (ONC) and observed a steep decline in the brown dwarf regime. Their census was focused on the low-mass range and they successfully fit a log-normal function to the obtained mass distribution. We use the best-fit parameters, namely the characteristic mass $m_c$ and the standard deviation $\sigma\, (\mathrm{log\,} m)$, from table~3 of \citet{2012ApJ...748...14D} to reproduce the log-normal fit to the mass distribution they derived by assuming a \citet{1998A&A...337..403B} evolutionary model. We then combine it with a Salpeter-like power-law tail, similar to what was done in \citet{2012ApJ...754...71K}. \citet{2013pss5.book..115K} argue that stars and brown dwarfs (BDs) must be represented by separate mass functions because the hypothesis that BDs form in the same manner as stars contradicts the observed binary properties of BDs. The \citet{2013pss5.book..115K} stellar system IMF \citep[taken from Fig.~25 in][]{2013pss5.book..115K} and BD IMF (short-dashed and long-dashed lines) result from random pairing of companions out of the canonical IMF \citep{2001MNRAS.322..231K} by assuming initial binary fractions of 100\% and 0\%, respectively. \citet{2021MNRAS.504.2557D} (dash-dot-dotted line) studied the role of environmental conditions on the form of the IMF by analysing the mass distribution of nine young clusters that differ in terms of the stellar density, number of massive stars and the Galactocentric distance. They observe no significant disparity between the mass distributions and find that the low-mass end of the IMF can be fitted by a log-normal distribution peaked at $0.32\pm0.02$ and $\sigma=0.47\pm0.02$ (in logarithmic scale).

The NOWIND SMD (top panel in Fig.~\ref{fig:imf_obs}) peaks at around $1\, \mathrm{M_\odot}$ while the the system IMFs of \citet{2005ASSL..327...41C} (dotted line) and \citet{2012ApJ...748...14D} (solid line) have peak masses of $\sim 0.25\, \mathrm{M_\odot}$ and $0.35\, \mathrm{M_\odot}$, respectively. In contrast, the broad peak of our OUTFLOW SMD (bottom panel in Fig.~\ref{fig:imf_obs}) is located at around $0.4$--$0.6\, \mathrm{M_\odot}$. Clearly, the introduction of the outflow feedback has resulted in a mass scale of the sink particles comparable to that of the observational models, although the peak mass is still higher by a factor of $\sim 2$. The slope of the power-law fit to the high-mass end of our SMD (OUTFLOW; $\Gamma=1.5\pm 0.3$) is slightly steeper than the Salpeter slope ($\Gamma=1.35$), but is well within the $1\sigma$ uncertainties, both from our fit and from observational estimates \citep{2013ApJ...762..123W,2014prpl.conf...53O}. The median stellar mass of our SMD is $ 0.5\pm0.1\, \mathrm{M_\odot}$ and the average mass is $0.9\pm0.1\, \mathrm{M_\odot}$ (at SFE = 5\%). We also calculate $M_{\mathrm{50}}$ which is, as defined in \citet{2012ApJ...754...71K}, the $50^{\mathrm{th}}$ percentile mass associated with a cumulative mass distribution, i.e., the sum of all the stellar masses lower than $M_{\mathrm{50}}$ is half the total mass of the distribution. We find $M_{\mathrm{50}}=1.5\, \mathrm{M_\odot}$ which is close to the value of $\sim 2\, \mathrm{M_\odot}$ for the observed IMFs. The ratio of the number of sink particles with sub-stellar masses ($M_{\mathrm{sink}} \leq 0.08\, \mathrm{M_\odot}$) to that of the sink particles with stellar masses ($0.15 < M_{\mathrm{sink}} \le 1.0\, \mathrm{M_\odot}$) is $43/225 = 0.19$, which is consistent with observations (where close binaries are unresolved) that approximately one BD is formed per every five late-type (sub-solar) stars \citep{2006AJ....132.2296A,2008ApJ...683L.183A,2007ApJ...671..767T,2011ApJ...726...27P,2013pss5.book..115K}.

\subsection{Comparisons with gravo-turbulent theoretical models of the IMF}

\begin{figure}
    \centering
    \includegraphics[width=\columnwidth]{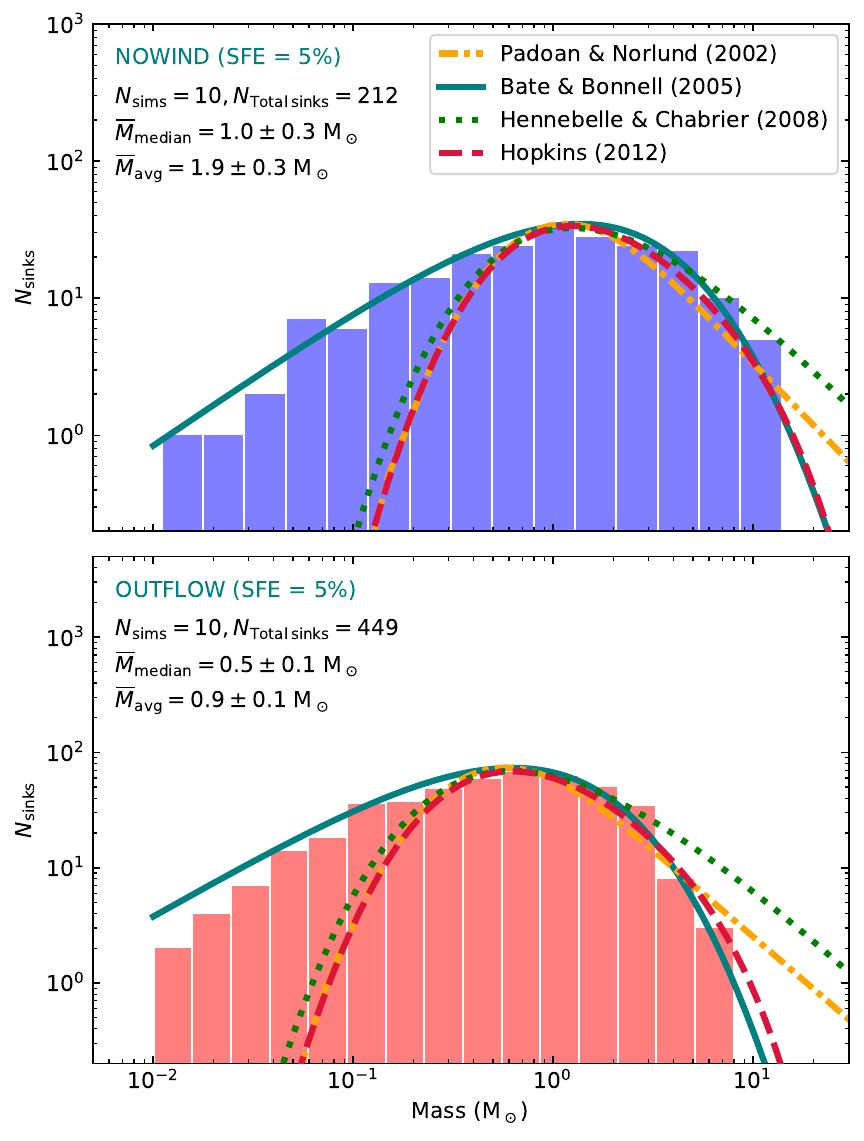}
    \caption{Top panel: Comparison between the sink mass distribution of the NOWIND model at SFE = 5\% with various theoretical models of the CMF/IMFs ($\mathcal{M}=5$, $\zeta=0.5$, $\beta=2$, $M_{\mathrm{J, 0}}\approx2$). The curves correspond to \citet{2002ApJ...576..870P} (dash-dotted), \citet{2005MNRAS.356.1201B} (solid), \citet{2008ApJ...684..395H} (dotted), and \citet{2012MNRAS.423.2037H} (dashed) CMF/IMFs. The \citet{2012MNRAS.423.2037H} model has been shifted to lower masses by a factor of 2 such that the position of their peak coincides with the peak mass bin of the SMD. Bottom panel: Same as the top panel, but for the OUTFLOW simulations. Here the gravo-turbulent theoretical models \citep{2002ApJ...576..870P,2008ApJ...684..395H,2012MNRAS.423.2037H} have been shifted to lower masses by a factor of 2 from their positions in the top panel in order to account for the effect of outflows (core-to-star efficiency $\epsilon=0.5$).} 
    \label{fig:imf_theo}
\end{figure}

In Fig.~\ref{fig:imf_theo}, we compare the NOWIND (top panel) and OUTFLOW (bottom panel) SMDs with the mass function (MF) predicted by different theoretical models of the CMF/IMF for parameters relevant to our simulation setup, e.g., the Mach number $\mathcal{M}=5$ and the turbulence driving parameter $\zeta=0.5$. The resulting functions from the gravo-turbulent theories \citep{2002ApJ...576..870P,2008ApJ...684..395H,2012MNRAS.423.2037H} are analogous to the mass distribution of cores (CMF) and the models are generally multiplied by a core-to-star efficiency factor $\epsilon$ (primarily to account for the impact of outflows) to facilitate comparison with the observational IMFs or the MF obtained in numerical studies. Therefore the comparison between the NOWIND SMD and the theoretical MFs is almost equivalent to a comparison between the OUTFLOW SMD and the theoretical MFs multiplied by an efficiency factor ($\epsilon\sim0.5$ in our study).

\subsubsection{The \citet{2002ApJ...576..870P} model}
According to the \citet{2002ApJ...576..870P} model, star-forming cores are the densest regions in the shocked layers of gas formed by supersonic turbulence. The model approximates the size of cores as the thickness of post-shock gas which in turn depends on the Mach number through the isothermal shock jump conditions. The scale dependence of Mach number \citep{1981MNRAS.194..809L} and the connection between the Larson relation and the velocity power spectrum result in a distribution of core masses for which the slope of the high-mass tail can be calculated from the velocity power spectral index $\beta$ in $P(k)\propto k^{-\beta}$ \citep{2021MNRAS.503.1138N}. The mass distribution of unstable cores is given by
\begin{equation}
    N(M)\, d\mathrm{log}M \propto M^{-\Gamma} \left[\int_{0}^{M} p(M_{\mathrm{J}})\, \mathrm{d}M_{\mathrm{J}}\right]\, d\mathrm{log}M,
\end{equation}
where $p(M_{\mathrm{J}})\, \mathrm{d}M_{\mathrm{J}}$ is the distribution of the Jeans mass and the integral gives the fraction of cores of mass $M$ that are Jeans unstable. For $\beta=2$, which is the typical one-dimensional power spectral index found for molecular clouds in observations and simulations \citep{2002A&A...390..307O,2004ApJ...615L..45H,2011ApJ...740..120R,2013MNRAS.436.1245F,2021NatAs...5..365F}, the high-mass slope of the IMF based on the MHD shock jump conditions is predicted to be \citep{2002ApJ...576..870P}
\begin{equation}
    \Gamma=3/(4-\beta)=1.5.
\end{equation}
The mass distribution at low masses is decided by the Jeans mass distribution, which in turn is determined by the probability density function (PDF) of the turbulent gas, which is approximately log-normal with a standard deviation defined by $\sigma_{\mathrm{s}}^2=\mathrm{ln}(1+b^2\mathcal{M}^2)$ \citep[see][]{1997MNRAS.288..145P,2008ApJ...688L..79F}. The peak of the distribution is influenced by the scale of the mean thermal Jeans mass $M_{\mathrm{J, 0}}$, which is $\sim 2$--$3\, \mathrm{M_\odot}$ in our simulations.

The \citet{2002ApJ...576..870P} model is represented by dash-dotted curves in both the top and bottom panels of Fig.~\ref{fig:imf_theo}. The mass distribution has been reproduced by using $\mathcal{M}=5$, $b=0.4$, $\beta=2$ and $M_{\mathrm{J, 0}}=2$ as the input parameters for their model. In the bottom panel, the theoretical models have been shifted to lower masses by a factor of 2 from their positions in the top panel in order to account for the impact of the outflow feedback.    

\subsubsection{The \citet{2008ApJ...684..395H} model}
\citet{2008ApJ...684..395H} apply the Press-Schechter formalism \citep{1974ApJ...187..425P} to study condensations in molecular clouds. In this context, the over-densities or density contrasts are created by supersonic turbulent motions. The spectrum of collapsing cores is obtained by statistically calculating the mass fraction of self-gravitating regions assuming a Gaussian field of density fluctuations. These regions are identified based on whether their mass exceeds the Jeans mass, also taking into account the support by turbulent pressure. The derived analytical function for the CMF/IMF consists of a log-normal component and a power-law component. The mass distribution has a power-law nature in the high-mass regime, but gradually flattens and then declines sharply at $M \ll (1+b^2\mathcal{M}^2)^{-3/2}\, M_{\mathrm{J, 0}}$. Using their expression for the slope of the power-law mass spectrum based on the turbulence power spectral index $\beta=2$, the prediction for the high-mass slope of the IMF is \citep{2008ApJ...684..395H} 
\begin{equation}
    \Gamma \approx (\beta+3)/2\beta=1.25.
\end{equation}
We mention that, according to \citet{2013ApJ...770..150H}, the high-mass slope becomes slightly steeper if the time-dependence is taken into account in the derivation of the IMF. We plot the \citet{2008ApJ...684..395H} CMF/IMF by using Eq.~44 in \citet{2008ApJ...684..395H} with input parameters as $\mathcal{M}=5$, $b=0.4$, $\beta=2$, $M_{\mathrm{J, 0}}=2$ and $\mathcal{M_*}=1.4$, which approximately represent our simulation configuration. The effective Mach number $\mathcal{M_*}$ corresponds to the velocity dispersion on the scale of the mean Jeans length. The model is depicted by dotted curves in Fig.~\ref{fig:imf_theo}. 

\subsubsection{The \citet{2012MNRAS.423.2037H} model}
\citet{2012MNRAS.423.2037H} employ the excursion set theory \citep{1991ApJ...379..440B} based on the principle of random walks to obtain the mass spectrum of cores (last crossing distribution) and giant molecular clouds (first crossing distribution). An important signature of the model is that, for a Gaussian distribution of density contrasts, the density variance at any given scale is not assumed but is inferred from the ISM properties. The form of the MF strongly depends on the Mach number at the injection scale of turbulent energy, characterised by the Mach number $\mathcal{M}_{h}$ on the galactic scale-height. Further, the theory resolves the \lq cloud-in-cloud\rq problem, i.e., the over-counting associated with a structure embedded in another structure of a larger scale. The MF transitions from a power law to a log-normal behaviour at the turnover mass $M_{\mathrm{sonic}}$, which is defined by the sonic scale $R_{\mathrm{sonic}}$, i.e., the scale at which the turbulent flow becomes subsonic \citep{2021NatAs...5..365F}. We use the Python code developed by \citet{2021MNRAS.503.1138N} to reproduce the \citet{2012MNRAS.423.2037H} mass function (dashed line in Fig.~\ref{fig:imf_theo}). We point out that here we define $\mathcal{M}_{h}=5$, which is the Mach number corresponding to the velocity dispersion on the driving scale of the turbulence ($L/2$) in the simulations. Our simulations do not have a characteristic scale height because of the periodic boundary conditions. \citet{2021MNRAS.503.1138N} show that such an ambiguity in the choice of $\mathcal{M}_{h}$ can significantly affect the results, particularly, the shape of the IMF as predicted in the \citet{2012MNRAS.423.2037H} model.

The mass functions of the above three theoretical models matches considerably well with our IMF in the high-mass regime and down to the low-mass range, but falls off exponentially towards the very low-mass range, faster than our SMD, i.e., the models do not predict any brown dwarfs based on the input parameters relevant to our simulations.

\subsection{Formation of sub-stellar objects}
\begin{figure*}
    \centering
    \includegraphics[width=\textwidth]{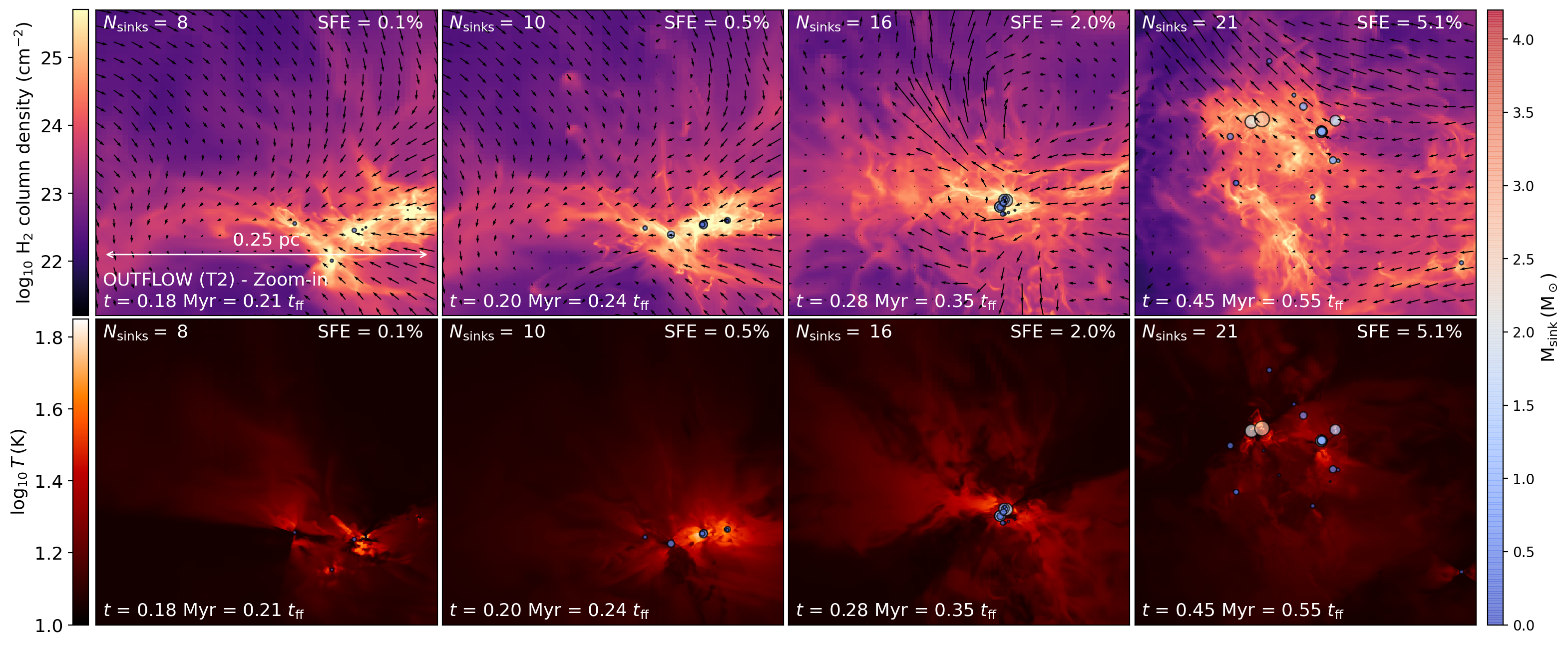}
    \caption{Time evolution (panels from left to right) of a clump in the T2 simulation (same simulation as the one considered in Fig.~\ref{fig:t2_compare}, but a different region of the cloud is studied here) of the OUTFLOW model illustrating the formation of sub-stellar objects. }
    \label{fig:t2_BD}
\end{figure*}

For the cloud properties adopted in our simulations, i.e., $\mathcal{M} = 5$, $\zeta=0.5$, $T = 10\, \mathrm{K}$, $n = \rho_{\circ}/(2.35\,m_{\mathrm{H}}) = 1.7 \times 10^3\, \mathrm{cm^{-3}}$, the three theoretical models of the IMF introduced above predict very few dense cores small enough to produce BDs \citep{2004ApJ...617..559P,2008ApJ...684..395H,2009ApJ...702.1428H}. In fact, we rarely observe sub-stellar objects forming in such small, marginally isolated over-densities in our simulations. Some of our very low-mass (VLM) sink particles (< $0.1\, \mathrm{M_\odot}$) were the ones that were ejected from high-order systems, which truncates their accretion \citep{2001AJ....122..432R,2002MNRAS.332L..65B}, and some formed as companions to other sinks that are slightly higher in mass. Most of the VLM objects in our simulations emerged in small clumps where a cluster of stars is formed. Fig.~\ref{fig:t2_BD} presents an example of the formation of VLM objects in a clump. The first snapshot shows sink particles forming at different locations of the clump. With time, they grow in mass and move towards the bottom of the potential well. When the stars are in close proximity to each other, multiple new objects are formed. These objects are likely to have formed due to fragmentation of the extended discs \citep{2002MNRAS.332L..65B,2007A&A...466..943G,2007MNRAS.382L..30S,2011ApJ...730...32S,2012MNRAS.423.1896R,2015ApJ...800...72T} of some of the initial sink particles, encouraged by dynamical interactions, e.g., star-disc or disc-disc collisions \citep{1998MNRAS.300.1189B,1998MNRAS.300.1214W,2010MNRAS.401..727S}. \citet{2010ApJ...717..577T} find that BDs can form by tidally induced fragmentation in extended discs due to close encounters with another star. Towards the end of the simulation, the cluster is dispersed because of gas removal from the clump. It is evident from the last time snapshot that in the process of clustering and eventual dispersal, at least 8~VLM objects were formed.

It is in the clustered regions, similar to the above example, that the influence of stellar luminosity (including that due to accretion) is paramount. Despite the fact that the heating of the disc by the central source is significantly reduced due to the shielding by dust particles, the disc is within the radius of influence of all the other stars because of the crowding. The increase in temperature due to the overlap of heating zones of all the stars would be enough to reduce the degree of fragmentation \citep{2020MNRAS.496.5201M}. Although outflows weaken the overall influence of stellar radiative heating feedback, the gas temperature would still be high enough to limit fragmentation, because of the contribution from multiple stars (see more examples of BD formation in Fig.~\ref{fig:t3_BD} and Fig.~\ref{fig:t4_BD} of Appendix A).

Since our results suggest that dynamical interactions are important for the formation of very low-mass objects, we also compare our simulation IMFs with the semi-analytical \citet{2005MNRAS.356.1201B} IMF model, which proposes that the IMF is controlled by the combination of accretion \citep{1997MNRAS.285..201B,2001MNRAS.323..785B} and stochastic dynamical ejection \citep{2001AJ....122..432R,2002MNRAS.332L..65B}.

\begin{table*}
	\caption{Calculated parameter values for the \citet{2005MNRAS.356.1201B} IMF model.}
	\label{tab:bbparams}
	\begin{tabular}{lccccc} 
	    \hline
		\hline
		 Model & $\overline{\dot M}_{\mathrm{acc}}\, [\mathrm{M_{\odot}}\, \mathrm{yr^{-1}}]$ & $\sigma_{\mathrm{acc}} (\mathrm{log\,} M)$ & $t_{\mathrm{eject}}\, [\mathrm{yr}] $  & $\mathrm{T_p}\, [\mathrm{yr}]$\\
        (1) & (2) & (3) & (4) & (5) \\
		\hline
		\hline 
		1. NOWIND  & $1.6 \times 10^{-5}$ & $0.29$ & $1.05 \times 10^5$ & $3.08 \times 10^5$\\
		2. OUTFLOW & $5.5 \times 10^{-6}$ & $0.26$ & $1.33 \times 10^5$ & $4.32 \times 10^5$\\
		
		\hline
	\end{tabular}
	\\
    \raggedright\textbf{Notes.} The values listed here are averages of the parameter values derived from the 10~simulations (realisations of the turbulence) for each simulation set, NOWIND and OUTFLOW. The \citet{2005MNRAS.356.1201B} IMF fits (solid curves in Fig.~\ref{fig:imf_theo}) have been obtained by using these parameter values.
\end{table*}

\subsubsection{The \citet{2005MNRAS.356.1201B} model}
\citet{2005MNRAS.356.1201B} IMF model assumes that all stellar and sub-stellar objects form with similar masses defined by the opacity limit for fragmentation. These objects then grow in mass at a constant rate until their accretion is terminated by dynamical ejection from dense gas region they formed in. The distribution of the accretion rate among the objects is assumed to be log-normal and the probability that an object is ejected at time $t$ is proportional to $\exp\left(-t/t_{\mathrm{eject}}\right)$, where $t_{\mathrm{eject}}$ is the characteristic ejection time-scale. The \citet{2005MNRAS.356.1201B} fits to our simulation IMFs can be obtained by calculating the following parameters: the mean accretion rate $\overline{\dot M}_{\mathrm{acc}}$, the standard deviation in the accretion rates $\sigma_{\mathrm{acc}} (\mathrm{log\,} M)$, the characteristic ejection time $t_{\mathrm{eject}}$, the minimum stellar mass $M_{\mathrm{min}}$, and the time period of the cluster formation $\mathrm{T_p}$. The quantity $\overline{\dot M}_{\mathrm{acc}}\,t_{\mathrm{eject}}$ gives approximately the peak mass of the IMF in the model. The dispersion in the accretion rates $\sigma_{\mathrm{acc}}(\mathrm{log\,} M)$ defines the high-mass and low-mass slopes, and the minimum stellar mass $M_{\mathrm{min}}$ sets the low-mass cut-off of the IMF. The time period $\mathrm{T_p}$ is the time elapsed between the formation of the first star and the end of the simulation. The calculated parameter values for each of the models are shown in Tab.~\ref{tab:bbparams}, which correspond to the averages of the parameter values obtained in the 10~simulations for each set (NOWIND and OUTFLOW). We set $M_{\mathrm{min}}= 0.01\, \mathrm{M_\odot}$ as the low-mass cut-off of the IMF fit for both the simulation models and substitute the above parameter values into equations~10--12 of \citet{2005MNRAS.356.1201B} to reproduce their fits for our simulation IMFs. The \citet{2005MNRAS.356.1201B} model is represented by solid curves in Fig.~\ref{fig:imf_theo}. The model matches well with our simulation IMFs, especially in the very low-mass regime, which was underestimated by the models based on turbulent fragmentation. This shows that stochastic dynamical ejections play an important role in determining the low-mass end of the IMF \citep[see also][]{2004MNRAS.347L..47B,2010MNRAS.405..401D,2011ApJ...743...98M,2014MNRAS.439..234M}. 

\subsection{Stellar kinematics}
\begin{figure}
    \centering
    \includegraphics[width=\columnwidth]{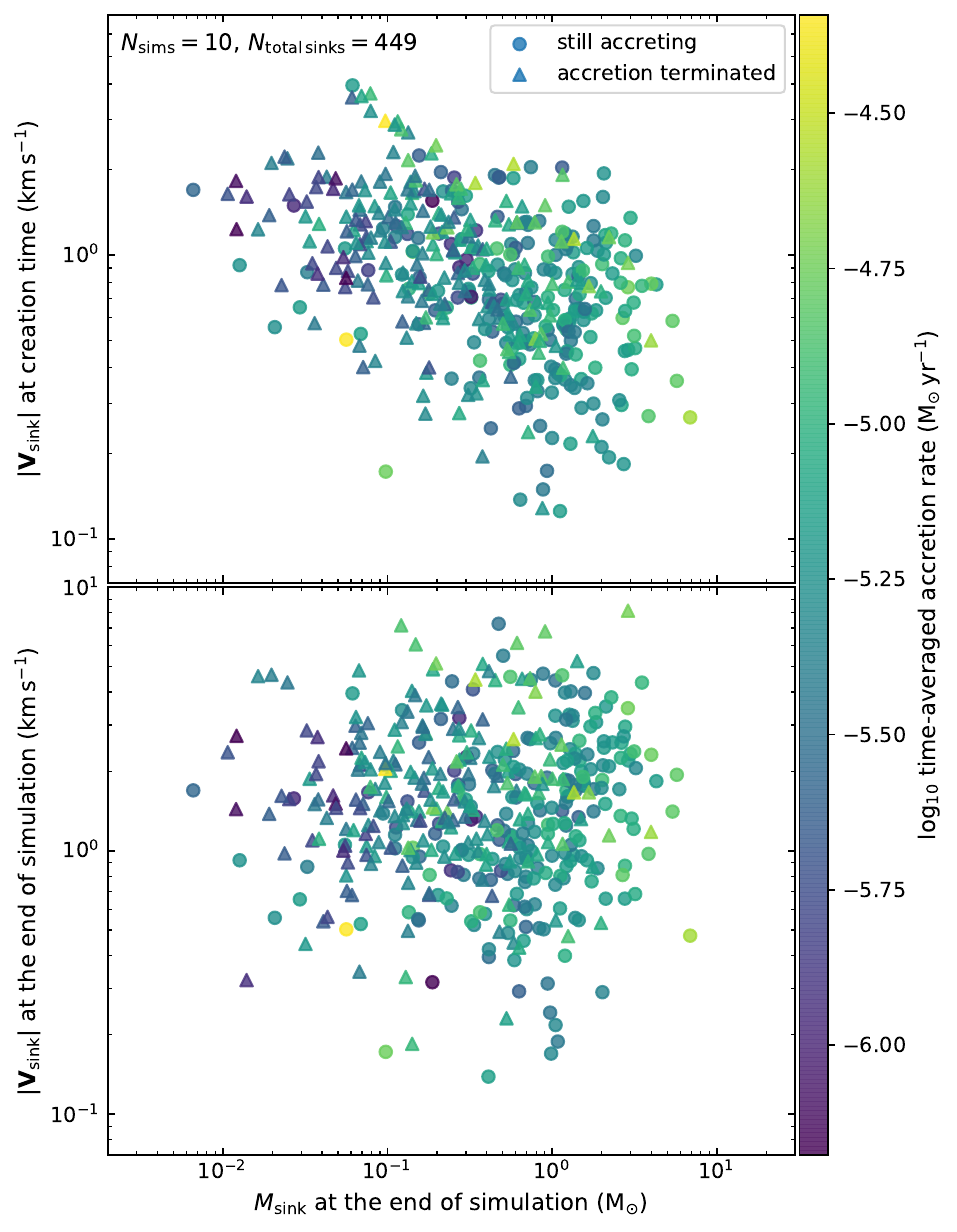}
    \caption{Top panel: velocity of the sink particles immediately after formation, as a function of their mass at SFE = 5\%. The circular markers represent the sink particles that are still accreting and the triangular markers represent the sinks that have stopped accreting when SFE = 5\% is reached. Each of the markers are colour-coded based on the time-averaged accretion rate of the sink particle (see right-hand colour scale). The data compilation is obtained from the ten OUTFLOW simulations. Bottom panel: same as top panel, but for the sink particle velocities at SFE = 5\%.} 
    \label{fig:velvsmass_tform}
\end{figure}

\begin{figure}
    \centering
    \includegraphics[width=\columnwidth]{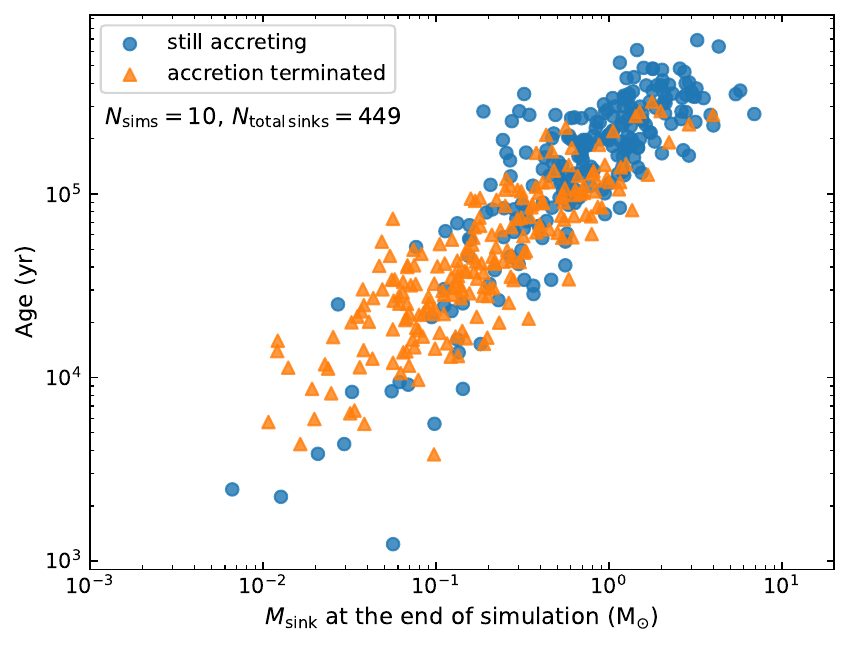}
    \caption{Age of the sink particles at the end of the corresponding simulation (SFE = 5\%) in the OUTFLOW model. The circular markers represent the sink particles that are still accreting and the triangular markers represent the sinks that have stopped accreting at SFE = 5\%. The age of the sink particles follows a roughly linear dependence on their final mass, similar to the simulations by \citet{2012MNRAS.419.3115B}.}  
    \label{fig:agevsmass}
\end{figure}

Fig.~\ref{fig:velvsmass_tform} shows the sink particle velocity $|\mathbf{V_{\mathrm{sink}}}|$ immediately after creation (top panel) and $|\mathbf{V_{\mathrm{sink}}}|$ at the end of the simulations, i.e., at SFE = 5\% (bottom panel) as functions of the their final mass (mass at SFE = 5\%) in the OUTFLOW model. The top panel seems to exhibit a trend of low-mass objects having a higher velocity than the high-mass objects, i.e., the velocity at the birth of the objects correlates (negatively) with their final mass to some extent, which is consistent with the findings of \citet{2008MNRAS.389.1556B}. However, there is no clear correlation between the velocity of the objects at the end of the simulation and their final mass. The apparent uncorrelated behaviour in the bottom panel is because of the dynamical evolution of the objects (clustering and eventual dispersal as discussed above), which rearrange the velocities such that some of the more massive objects ($> 0.5\, \mathrm{M_\odot}$) attain higher velocities at later times, and therefore, the trend of the formation $|\mathbf{V_{\mathrm{sink}}}|$ with mass seen in the top panel, is lost at later times. 

Fig.~\ref{fig:agevsmass} shows the age of all the sink particles in our simulations, i.e., the time between their formation and the end of the simulation as a function of their final mass. The age of the sink particles follows a linear relationship with the final mass. The sink particles with masses in the BD regime are the youngest and the high-mass sink particles are the oldest. \citet{2012MNRAS.419.3115B} also observed such a linear trend in their simulations.

Figs.~\ref{fig:velvsmass_tform} and~\ref{fig:agevsmass} show that (1) the sink particles that reach high masses are the first to form in the simulations and have low velocities at the time of formation. Their velocities are modified due to the subsequent interaction with the gas and other sink particles. (2) A large number of the very low-mass objects form towards the end of the simulations, which have high velocities at the time of formation.

We find that the 1D velocity dispersion of the sink particles (at SFE = 5\%) varies between the simulations and they lie in the range $0.65$--$1.74\,\mathrm{km\, s^{-1}}$ with a mean value of $1.1\pm0.3\, \mathrm{km\, s^{-1}}$. The velocity dispersion measured by observational surveys in the ONC is around $2\, \mathrm{km\, s^{-1}}$ \citep{1988AJ.....95.1755J,2009ApJ...697.1103T,2019AJ....157..109K,2021arXiv210505871T}. In the inner core region of the ONC, \citet{2021arXiv210505871T} derived a slightly higher 1D velocity dispersion of $\sim 2.5\, \mathrm{km\, s^{-1}}$. \citet{2019ApJ...870...32K} measured 1D velocity dispersions in the range $0.8$--$2.8\, \mathrm{km\, s^{-1}}$ for a sample of different young stellar clusters and associations.

\section{Multiplicity and angular momentum}
\label{sec:Multi}

\subsection{Multiplicity fraction}
We identify the multiple systems in our simulations by following the algorithm used in \citet{2009MNRAS.392..590B}. In the list of individual sink particles formed in a simulation, or equivalently a list of single objects, we find the closest pair of objects that are gravitationally bound to each other. These two objects are replaced by a binary object having the mass, position and velocity equal to the system mass, centre-of-mass position and velocity, respectively. In the resulting list, we again find the closest pair of bound objects. In case the pair consists of a binary object and a single object, then the new object will be a triple. The process of removing the closest bound pair and replacing it with a higher-order object is repeated until none of the objects in the list are bound to one another or the only possible gravitationally bound pair selection will result in a quintuple. We do not consider systems of order higher than quadruples, because most high-order multiple systems are dynamically unstable and are expected to decay with further evolution of the cloud. 

The algorithm effectively converts a list of individual sink particles into a list of single, binary, triple and quadruple systems or objects, with none being a subset of a higher-order system. For example, none of the objects characterised as binaries is a part of a triple or quadruple object. The multiplicity fraction in any mass range can be obtained by calculating the ratio of the number of multiple systems to the total number of systems whose primary sink particle lie within the given mass range. The multiplicity fraction ($mf$) is defined as   

\begin{equation}
    mf =  \frac{B + T + Q}{S + B + T + Q}
\end{equation}
where $S, B, T$ and $Q$ represent the number of singles, binaries, triples and quadruples, respectively, whose primary sink's mass is within the range in which $mf$ is to be calculated.

\begin{figure}
    \centering
    \includegraphics[width=\columnwidth]{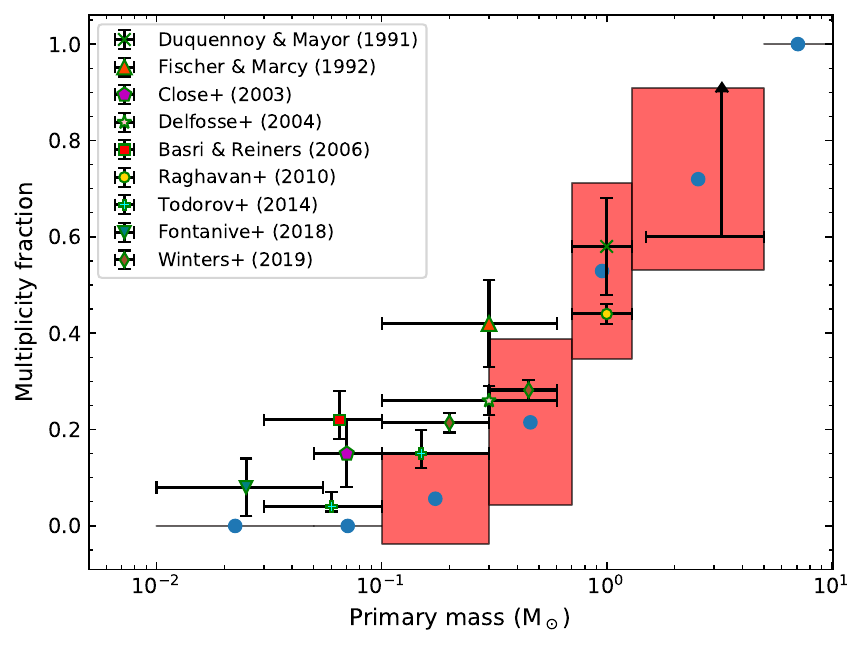}
    \caption{Multiplicity fraction ($mf$) as a function of primary mass. The circular markers represent the average $mf$, obtained from all our 10~simulations, in the mass interval represented by the width of the patch enclosing the marker. The height of the patch shows the standard deviation of $mf$ across the 10~simulations. The centre of the crosses represent the value of $mf$ obtained in different observational surveys, with the horizontal and vertical error bars corresponding to the mass range of study and the uncertainties, respectively. The observational data are (from left to right), from \citet{2018MNRAS.479.2702F}, \citet{2014ApJ...788...40T}, \citet{2006AJ....132..663B}, \citet{2003ApJ...587..407C}, \citet{2014ApJ...788...40T}, \citet{2019AJ....157..216W} (not corrected for undetected companions), \citet{2004ASPC..318..166D}, \citet{1992ApJ...396..178F}, \citet{2010ApJS..190....1R} and \citet{1991A&A...248..485D}. The multiplicity fraction of high-mass stars is relatively poorly understood. The lower limit of $mf$ in the mass range of $1.5$--$5\, \mathrm{M_\odot}$ is $\sim$ 0.5--0.6 \citep{2012MNRAS.424.1925C,2013ARA&A..51..269D}. Massive stars are expected to have $mf\sim1$ \citep{2009AJ....137.3358M,2011IAUS..272..474S,2017A&A...599L...9S,2020SSRv..216...70L}.}
    \label{fig:Multiplicityfraction}
\end{figure}

Fig.~\ref{fig:Multiplicityfraction} depicts the multiplicity fraction in different primary mass intervals \citep[also done in][]{2012MNRAS.419.3115B,2012ApJ...754...71K,2018MNRAS.476..771C,2020MNRAS.497..336S} at SFE = 5\% in the case of the OUTFLOW model, which is the more realistic model. The mass ranges are selected similar to those chosen in the observational studies so as to allow for a direct comparison. We can immediately notice that the multiplicity fraction evolves as an increasing function of the primary mass, which is the general understanding. Our $mf$ values also compare well with those of the observations, except that we are underestimating the value in the very low-mass star (VLMS) and BD regime.

Since our highest attainable spatial resolution is 100~AU, we do not resolve all of the low-order multiple systems. Therefore, some of the sink particles may be representing binaries by themselves, or rarely, triple systems. However, the multiplicity fraction is very robust numerically. Even if a sink particle can be further fragmented into a binary or a triple stellar system, $mf$ increases only if the sink particle is a single. The value remains unchanged if the sink particle is part of a multiple system, i.e., a member of a binary, triple or quadruple object. For example, if one of the sinks belonging to a triple object is a binary by itself, then $T\, \mathrm{and}\, Q$ become $T-1\, \mathrm{and}\, Q+1$, respectively, which leaves $mf$ unchanged. Further, considering the fact that the frequency of singles decreases and the average separation of binaries increases with increasing primary mass \citep{2007ApJ...663..394K,2007ApJ...662..413K,2012ARA&A..50...65L}, the mass range that will be mainly affected is the low-mass end, which explains the underestimation of $mf$ in the sub-solar regime. If a few of the sink particles in the mass range $0.01$--$0.1\,\mathrm{M_\odot}$ were actually binaries if we had higher numerical resolution, then the increase in the multiplicity fraction would balance the underestimation in the VLMS or BD regime, which would bring the simulations closer to the observations in the sub-solar mass regime.

\begin{figure}
    \centering
    \includegraphics[width=\columnwidth]{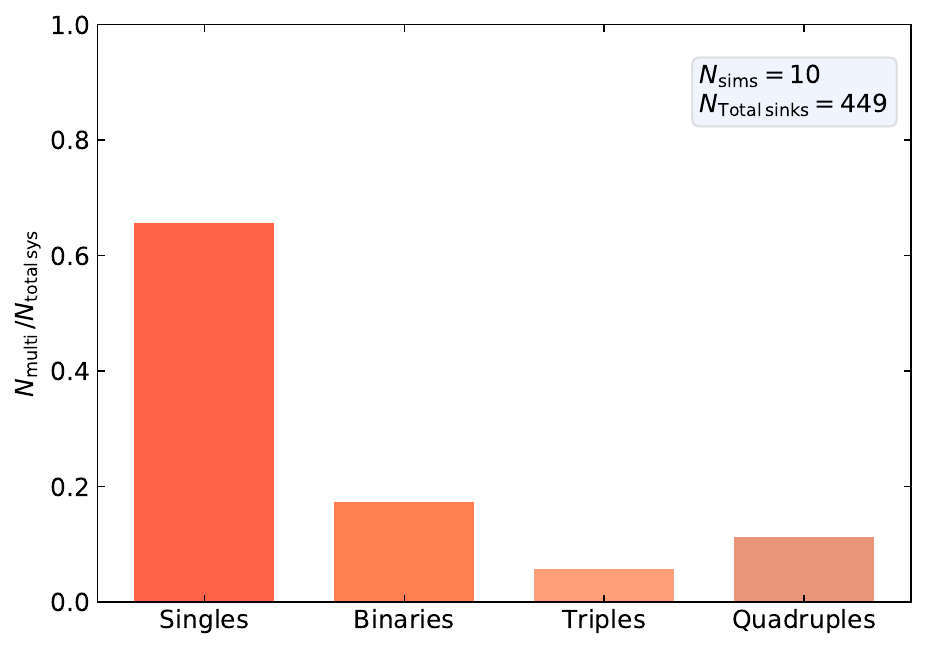}
    \caption{Fraction of single stars and multiple systems (binaries, triples, and quadruples), using the same data as for Fig.~\ref{fig:Multiplicityfraction}. The fraction of single objects is 0.6--0.7 in our simulations, which agrees well with the estimates of the single-star fraction of observed stars by \citet{2006ApJ...640L..63L}.}
    \label{fig:multiplicityplot}
\end{figure}

Fig.~\ref{fig:multiplicityplot} shows the fraction of singles, binaries, triples and quadruples at SFE = 5\%. The fraction of singles is 0.66 and is the highest, i.e., most of the sink particles that formed in our simulations are not part of a higher-order multiple system (the fraction of single objects for the NOWIND model is quoted in Tab.~\ref{tab:sims} for reference). \citet{2021MNRAS.500.3594R} also find that the fraction of singles is around $0.6$ in their simulations of turbulent dense cores. Adopting the multiplicity fraction derived by \citet{2004ASPC..318..166D} for M-dwarfs and that by \citet{1991A&A...248..485D} for stars earlier than M-dwarfs, \citet{2006ApJ...640L..63L} estimated the single-star fraction (SSF) to be 0.66 by assuming a \citet{2002ApJ...573..366M} form of the IMF and 0.67 for a \citet{1979ApJS...41..513M} IMF. Our value of the fraction of single objects agrees with their estimates, but the exact match between the values is coincidental. Observational investigations do not well constrain the IMF or the multiplicity fraction of different spectral types. Therefore, the SSF estimate of \citet{2006ApJ...640L..63L} would vary depending on the study they choose to derive the multiplicity fraction and the number fraction of M-dwarfs. Further, \citet{2006ApJ...640L..63L} only considered stars of spectral type M and earlier, whereas we also include the very low-mass stars and brown dwarfs. The actual value of SSF could be slightly lower than what we calculated because of the possibility of unresolved binaries in our simulations. It is also plausible that we are somewhat underestimating the fraction because, with further evolution of the cloud, some of the higher-order systems may decay and lead to an increase in the number of single stars.

\subsection{Companion separation distribution}
The binary separation distribution of solar-type main-sequence stars resembles a Gaussian function with a peak at $\sim\!45\, \mathrm{AU}$ and a standard deviation of 1.5 in the logarithm of the separations according to the observational survey by \citet{2010ApJS..190....1R} \citep[see also][]{1991A&A...248..485D}. \citet{2019AJ....157..216W} fit a Gaussian to their projected separation distribution of M-dwarfs with a peak at 20 AU and a value of 1.16 for the standard deviation of the log-separation. However, the separation distribution of younger systems are shown to differ from the separation distribution of the field stars \citep{2002AJ....123.1570P,2004A&A...427..651D,2008AJ....135.2526C,2013ApJ...768..110C,2016ApJ...818...73T}. The VANDAM survey \citep{2016ApJ...818...73T} studied a sample of Class~0 and Class~I protostars in the Perseus molecular cloud and obtained a double-peaked distribution with a peak at $\sim\!75\, \mathrm{AU}$ and a secondary peak at wide separations (> 1000 AU), although the secondary peak was absent for their sample of only Class~I protostars. There have been numerous numerical works in the past that study the formation of multiple systems and analyse their separation distribution \citep[e.g.,][]{2009MNRAS.392..590B,2010ApJ...725.1485O,2012MNRAS.419.3115B,2019ApJ...887..232L}.

\begin{figure}
    \centering
    \includegraphics[width=\columnwidth]{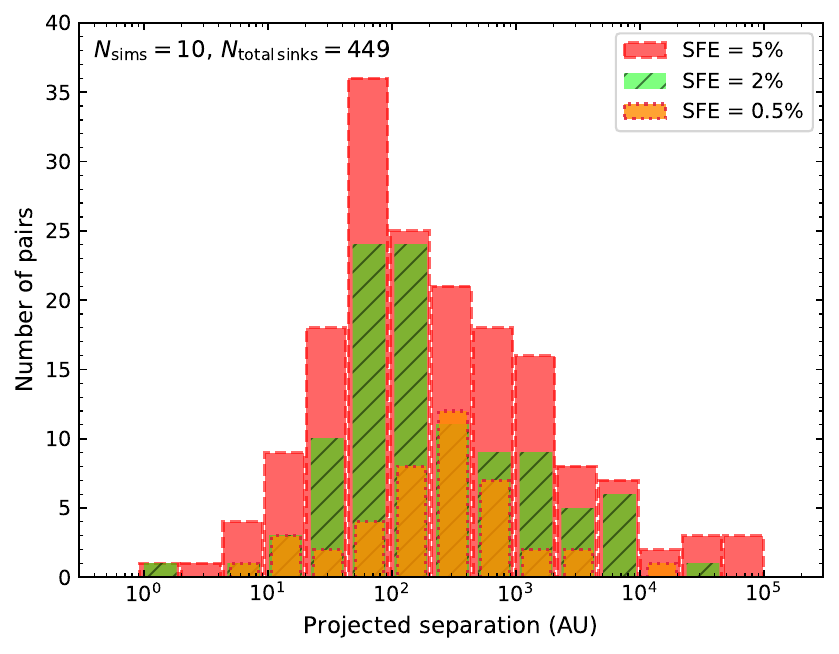}
    \caption{The projected separation distribution of the multiple systems in the ten OUTFLOW simulations at SFE = 0.5\% (orange histogram with dotted edges), 2\% (green hatched histogram) and 5\% (red histogram with dashed edges). The projected distance presented here is the average of the projections of the separation in the three Cartesian planes. All distances are measured from the primary object. A binary provides a single separation, a triple provides two, and a quadruple contributes three separations.} 
    \label{fig:sep_dis}
\end{figure}

The Class~0 lifetime of approximately $10^4$--$10^5\, \mathrm{yr}$ and Class~I lifetime of a few $10^5 \mathrm{yr}$ \citep{2006MNRAS.368..435F,2014prpl.conf..195D} are relatable to the age of our sink particles (see Fig.~\ref{fig:agevsmass}). Fig.~\ref{fig:sep_dis} shows the projected separation distribution of the multiple systems in our OUTFLOW simulations. In each of the systems, the separation distances are measured from the primary star. For example, a triple system will contribute two separation pairs to the distribution and a quadruple system contributes three pairs. The separation distribution at SFE = 0.5\%, 2\% and 5\% are shown by the orange (dotted edges), green (hatched), and red (dashed edges) histogram, respectively. The peak of the distribution at SFE = 0.5\% lies at around 300--400~AU. The statistical significance of the distribution at SFE = 0.5\% is low since the number of sink particles that form within that time is small compared to the total number at SFE = 5\%. The peak of the distribution shifts to shorter separations by the time SFE = 2\% is reached and the peak is at an even shorter separation at SFE = 5\%, indicating an evolution of systems to closer separations due to dynamical interactions \citep{2001AJ....122..432R,2002MNRAS.336..705B,2005A&A...439..565G,2010ApJ...725.1485O}. By the end of the simulations (at SFE = 5\%), there is an appreciable number of wide separations (>~1000~AU), even though separation distances of less than 500~AU are more common. For the distribution at SFE = 5\%, the standard deviation of the logarithm of the projected separations is 0.9, i.e., the dispersion is close to an order of magnitude. The number of short separations is only a lower limit since fragmentation on small scales (at <~$\sim\!200$~AU) does not occur in our simulations. Our sink particle formation criterion does not allow the formation of new sink particles within the accretion radius (250~AU) of an existing sink particle. All the pairs in the distribution with separation less than 250~AU correspond to the pairs that formed at a greater separation initially, and moved closer to one another at later times. At SFE = 5\%, the separation range 40--100~AU has the highest frequency of pairs. The comparison of the peak of our distribution with the observational results should be done with caution, because the gas dynamics on very small scales is poorly modelled in our simulations due to the limited resolution, and also the gravitational potential is softened within the accretion radius. Therefore, a resolution study would be required to find the converged location of the peak of the separation distribution.

\subsection{Specific angular momentum of dense cores and stars}
\label{sec:angmom}
\begin{figure}
    \centering
    \includegraphics[width=\columnwidth]{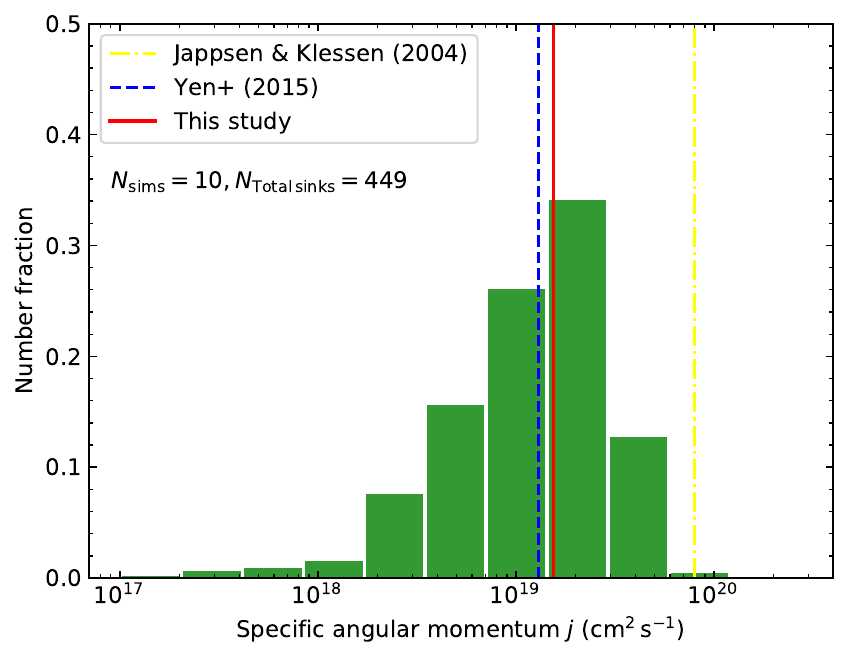}
    \caption{Specific angular momentum $j$ of the sink particles ($r_{\mathrm{sink}}= 250\, \mathrm{AU}$) from the OUTFLOW simulations (green histogram), with the solid line marking the mean value of $j$. The dashed line shows the $j$ value measured for the class~0 protostar B335 at $\sim\!$~180~AU by \citet{2015ApJ...812..129Y}, and the dash-dotted line denotes the mean value of $j$ obtained in the numerical simulations of \citet{2004A&A...423....1J} with a sink particle radius of 560~AU.}
    \label{fig:specific_angm}
\end{figure}

Previous studies estimate the specific angular momentum ($j$) of dense molecular cloud cores (diameter $\sim\!0.1\,$pc) to be greater than $10^{21}\, \mathrm{cm^2\, s^{-1}}$ \citep{1993ApJ...406..528G,2000ApJ...543..822B,2002ApJ...572..238C}. The range of $j$ value of class~0/I envelopes and binary systems is $10^{17}$--$10^{21}\, \mathrm{cm^2\, s^{-1}}$ \citep{1992ASPC...32...41S,1997ApJ...488..317O,2015ApJ...799..193Y}, whereas that of T-Tauri stars is $10^{16}$--$10^{17}\, \mathrm{cm^2\, s^{-1}}$ \citep{1986ApJ...309..275H}. The angular momentum of the sink particles (spin) in our simulations (see \S\ref{sec:sink}) can be directly obtained from the simulation output. We plot the specific angular momentum distribution of all the sink particles that formed in the 10~simulations of the OUTFLOW model in Fig.~\ref{fig:specific_angm}. The range of specific angular momentum of the sinks ($\sim\!10^{17}$--$10^{20}\, \mathrm{cm^2\, s^{-1}}$) spans the regime of protostellar envelopes and binaries, with a few sinks having $j$ values typical of T-Tauri stars. The average specific angular momentum of all the sinks is $j_{\mathrm{mean}} = 1.5 \times 10^{19}\, \mathrm{cm^2\, s^{-1}}$. \citet{2015ApJ...812..129Y} report the $j$ value of the class~0 protostar B335 to be $1.3 \times 10^{19}\, \mathrm{cm^2\, s^{-1}}$ at a scale of $\sim\!$~180~AU, which agrees well with the $j_{\mathrm{mean}}$ from the sink particles in our simulations (having an accretion radius of 250~AU). \citet{2020A&A...637A..92G} infer that the $j$ value of class~0 protostellar envelopes is relatively constant, at around $10^{20}\, \mathrm{cm^2\, s^{-1}}$, from a scale of $\sim\!$~1600~AU to 50~AU. \citet{2004A&A...423....1J} performed hydrodynamic simulations of the collapse of supersonic turbulent clouds and calculated $j_{\mathrm{mean}} = 8 \times 10^{19}\, \mathrm{cm^2\, s^{-1}}$ for their sink particles with a radius of 560~AU. We note that $j_{\mathrm{mean}}=1.2 \times 10^{20}\, \mathrm{cm^2\, s^{-1}}$ for the NOWIND model, which is almost an order of magnitude higher than that for the OUTFLOW model. This is because, based on observational and numerical works, our SGS outflow module removes 90\% of the accreted angular momentum from the sink particles and re-distributes it to the jet/outflow components (see \S\ref{sec:outflowfeedback}).

\begin{figure}
    \centering
    \includegraphics[width=\columnwidth]{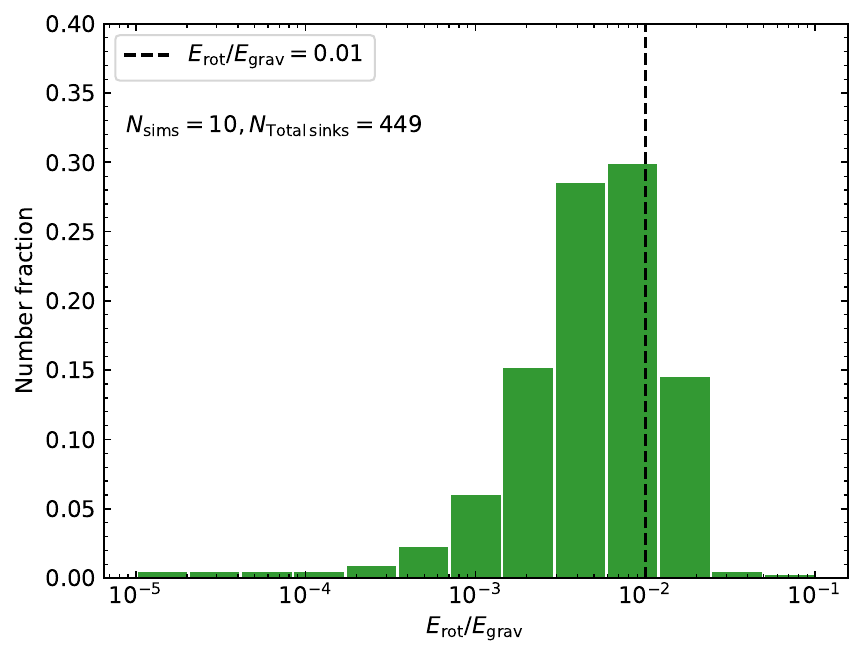}
    \caption{The distribution of the ratio of rotational to gravitational energy ($E_\mathrm{rot}/E_\mathrm{grav}$) of the sink particles formed in our simulations. The black dashed line marks $E_\mathrm{rot}/E_\mathrm{grav}=0.01$, above which fragmentation into binaries is likely \citep{1999ApJ...520..744B,2004A&A...423....1J}. Thus, only a relatively small fraction of our sink particles ($92/449=20\%$) would have likely fragmented further, if we had higher numerical resolution.}
    \label{fig:Erot}
\end{figure}

Fig.~\ref{fig:Erot} presents the distribution of the ratio of rotational to gravitational energy $E_\mathrm{rot}/E_\mathrm{grav}$ of the sink particles. We calculate $E_\mathrm{rot}/E_\mathrm{grav}$ by assuming solid-body rotation of a uniform density sphere,
\begin{equation}
    \frac{E_\mathrm{rot}}{E_\mathrm{grav}} = \frac{(1/2)I\omega^2}{(3/5)GM_{\mathrm{sink}}^2/r_\mathrm{sink}},
\end{equation}
where $I=(2/5)M_{\mathrm{sink}}r_\mathrm{sink}^2$ is the moment of inertia, $\omega=jM_{\mathrm{sink}}/I$ is the angular velocity, and $G$ is the gravitational constant.

We can roughly estimate the number of sink particles that represent unresolved binaries by looking at their $E_\mathrm{rot}/E_\mathrm{grav}$ values. The sink particles with $E_\mathrm{rot}/E_\mathrm{grav} < 0.01$ tend to be stable against fragmentation while those with $E_\mathrm{rot}/E_\mathrm{grav} > 0.01$ are likely to fragment into binaries \citep{1999ApJ...520..744B,2004A&A...423....1J}. We observe that, out of the total 449~sink particles from the 10~simulations, 92~sinks have $E_\mathrm{rot}/E_\mathrm{grav} > 0.01$. However, the multiplicity fraction would only be affected by the fragmentation of the stellar objects that are not already a part of a higher-order system (i.e., single sink particles). For example, out of the 449~total, there are only 7~sink particles with masses less than $0.1\, \mathrm{M_\odot}$ that are simultaneously single and have $E_\mathrm{rot}/E_\mathrm{grav} > 0.01$. Since the number of singles decreases with increasing primary mass, we expect that the $mf$ estimates discussed above would not deviate much from the current value in the higher-mass range.

\section{Caveats}
\label{sec:discussions}
\subsection{Numerical resolution}
Our sink mass distribution matches well with the form of the observed system IMFs. Despite the overall good agreement, higher numerical resolution might lead to a slight increase in fragmentation (c.f.~Fig.~\ref{fig:Erot}), and thus, may result in a mass distribution that matches the individual-star IMF or may produce the observed IMF peak mass better (currently overestimated by factor $\sim2$; c.f.~Fig.~\ref{fig:imf_obs}). Here our sink particles have radii of $r_{\mathrm{sink}}= 250\, \mathrm{AU}$, and the high computational cost of these simulations currently prevents us from going higher in numerical resolution, because at the same time, we want to produce a statistically converged sample of stars (here built up from 10 independent simulations with different random seeds of the turbulence). Recent studies by \citet{2019ApJ...883..140H} and \citet{2020MNRAS.492.4727C} suggest that the origin of the peak of the IMF may lie in the tidal screening of the first hydrostatic core or the Larson core. The characteristic mass of the IMF would then be determined by the typical mass within the radius in which the tidal force by the Larson core prevents formation of any fragments, in which case a resolution of $\sim\!$~10--20~AU would be necessary to obtain the peak mass. The multiplicity of stars can also be affected by the limitation in resolution. The current simulations do not resolve all the stars, particularly, the close binaries, which is why we somewhat underestimate the $mf$ in the sub-solar range (c.f.~Fig.~\ref{fig:Multiplicityfraction}). Very low-mass binaries generally have separations of $<\, 20\, \mathrm{AU}$, and therefore, sufficiently high-resolution simulations would be required to resolve such binaries.  Nevertheless, we believe that the multiplicity fraction estimates in the mass range $>\,0.1\, \mathrm{M_\odot}$ are relatively robust as discussed above. It is reassuring that state-of-the-art simulations are now being developed that can simultaneously follow the collapse from the scales of GMCs down to the disc scales and also include all the primary mechanisms involved in star cluster formation \citep{2021MNRAS.tmp.1371G}. 

\subsection{Use of sub-grid models}
The sub-grid models that we use for incorporating the effects of jets and outflows \citep{2014ApJ...790..128F} and stellar radiative heating \citep{2020MNRAS.496.5201M} have been calibrated and well tested. However, they are dependent on parameters that are based on previous observational and numerical studies. \citet{2021MNRAS.502.3646G} find that the characteristic mass of the IMF somewhat depends on the choice of the parameter that decides the amount of momentum injection by their outflow model. Therefore, variations in the mass-loss factor $f_\mathrm{m}$ or the normalisation jet speed ($100\, \mathrm{km\, s^{-1}}$) used to define the outflow velocity profile might cause the peak of our sink mass distribution to deviate from the current position. In addition, the fact that the polar heating model we use calculates the extinction of stellar radiation by assuming a fixed dust/disc density distribution can cause discrepancies. In order to accurately model jets or the extinction of stellar radiation by dust, the inner regions of the accretion disc have to be resolved. Achieving such high resolution generally comes at the expense of statistical significance because only one or a few simulations can be performed or the outer scale of the simulation box has to be reduced (smaller cloud size) to reduce the computational cost. Therefore, employing sub-grid models is necessary to obtain statistically conclusive and quantitatively accurate results in a cost-efficient way.

\section{Conclusions}
\label{sec:conclude}
We carry out an array of simulations of the collapse of turbulent, magnetized molecular clouds including stellar heating and mechanical feedback. We investigate the impact of jets/outflows on the IMF and the evolution of different dynamical quantities by comparing 10~simulations with different turbulence seeds incorporating outflow feedback with another set of 10~simulations in the absence of the outflow feedback. We observe that the outflow feedback reduces the star formation rate by a factor of $\sim2$ and increases the number of stars formed (see details in Tab.~\ref{tab:sims}). Jets/Outflows disturb the direct accretion flow onto a star and lower the efficiency of stellar heating (as a result of the mass loss) in suppressing fragmentation, and thereby promote star formation in nearby over-dense regions. We find that including outflow feedback does not generally affect the overall shape of the IMF, but results in a shift of the IMF to lower masses by a factor of \mbox{$2.0\pm0.2$}.

We find that the IMF obtained from our simulations broadly agrees with different functional forms of the observational IMF, although the peak mass is higher by a factor of $\sim2$ (see discussion in Sec.~\ref{sec:discussions}). We also show that three different theoretical models of the IMF based on turbulent fragmentation \citep{2002ApJ...576..870P,2008ApJ...684..395H,2012MNRAS.423.2037H} broadly predict the shape of our IMF in the high-mass and low-mass range, but underestimate the number of very low-mass objects ($< 0.1\, \mathrm{M_\odot}$). Indeed, the fraction of sub-stellar objects produced in our simulations agrees well with the observed fraction. Therefore, our current set of simulations suggests that some modification of the gravo-turbulent theoretical models is required to account for the population of very low-mass stars \citep[see also][]{2015ApJ...800...72T}. 

We also compare our simulation IMF with the \citet{2005MNRAS.356.1201B} model, which is based on accretion and dynamical ejections, and see that it matches well with our simulation IMF. Further, the \citet{2005MNRAS.356.1201B} model reflects our very low-mass range better than the gravo-turbulent models, indicating the importance of dynamical ejections in the formation of sub-stellar objects. It is interesting to see that the IMF can be reproduced directly on the basis of the accretion and ejection history of the young stars. The \citet{2005MNRAS.356.1201B} model only uses information about the stars and their evolution inside the cluster. On the other hand, the gravo-turbulent models only use the gas properties of the cloud to derive an IMF, not taking into account any information about the dynamical evolution of the stars when they have already formed. Therefore, the nature of the two classes of model (relying primarily on stellar versus gas properties) is fundamentally different. For example, \citet{2021MNRAS.503.1138N} find that the high-mass tail of the IMF is significantly influenced by the velocity power spectrum of the turbulence in the parent molecular cloud. Therefore, we suggest that both turbulent gas properties and the accretion and ejection history of the young stars play key roles in controlling the IMF, with a tendency of the high-mass end being controlled by turbulent gas properties, and the low-mass end being controlled by dynamical ejections and radiation feedback. 

In our simulations, the velocity of the protostars at their birth correlates with their final mass, such that the objects with masses in the very low-mass regime have high velocities and the high-mass stars have low velocities at the time of their formation, consistent with the simulations of \citet{2008MNRAS.389.1556B}. However, dynamical interactions during their lifetime modify the velocities such that no correlation remains by the end of the simulations. The 1D velocity dispersion of the protostars at the end of the simulations is $1.1\pm0.3\, \mathrm{km\, s^{-1}}$.

We find that the multiplicity fraction of the stellar systems in our simulations is an increasing function of the primary mass, consistent with other numerical studies and with observational surveys. Our multiplicity fractions compare well with observational estimates for different spectral types, although we are slightly underestimating the multiplicity fraction in the very low-mass range. This underestimation is mainly because our simulations do not fully resolve very close binaries. We see that, as more of the cloud gas gets converted into stars and the stellar density increases with time, the peak of the companion separation distribution of the multiple systems shifts to shorter separations, implying an orbital decay (hardening) of the systems \citep{2001AJ....122..432R,2002MNRAS.336..705B,2005A&A...439..565G,2010ApJ...725.1485O}, and therefore highlighting again the importance of dynamical interactions of the young stars.

The range of the specific angular momentum ($j$) of our sink particles can be directly compared to the $j$ values in protostellar envelopes and binaries. We find that the average specific angular momentum of the sink particles, $j_{\mathrm{mean}} = 1.5 \times 10^{19}\, \mathrm{cm^2\, s^{-1}}$, matches the value of specific angular momentum from observational measurements at a scale similar to the size of our sink particles.

\section*{Acknowledgements}
We thank the anonymous reviewer for their comments and valuable suggestions, which helped to improve the paper. We thank Pavel Kroupa for valuable discussions regarding the system IMF. We further thank Piyush Sharda for helpful comments on the multiplicity algorithm and Donghee Nam for assistance with the python code for reproducing the Hopkins IMF. C.F.~acknowledges funding provided by the Australian Research Council (Future Fellowship FT180100495), and the Australia-Germany Joint Research Cooperation Scheme (UA-DAAD). We further acknowledge high-performance computing resources provided by the Leibniz Rechenzentrum and the Gauss Centre for Supercomputing (grants~pr32lo, pr48pi and GCS Large-scale project~10391), the Australian National Computational Infrastructure (grant~ek9) in the framework of the National Computational Merit Allocation Scheme and the ANU Merit Allocation Scheme. The simulation software FLASH was in part developed by the DOE-supported Flash Center for Computational Science at the University of Chicago.

\section*{Data Availability}
The data used in this article is available upon request to the authors.



\bibliographystyle{mnras}
\bibliography{Bibliography} 




\appendix
\section{Column density and temperature maps of simulations with different random seeds for the turbulence}
\begin{figure*}
    \centering
    \includegraphics[width=\textwidth]{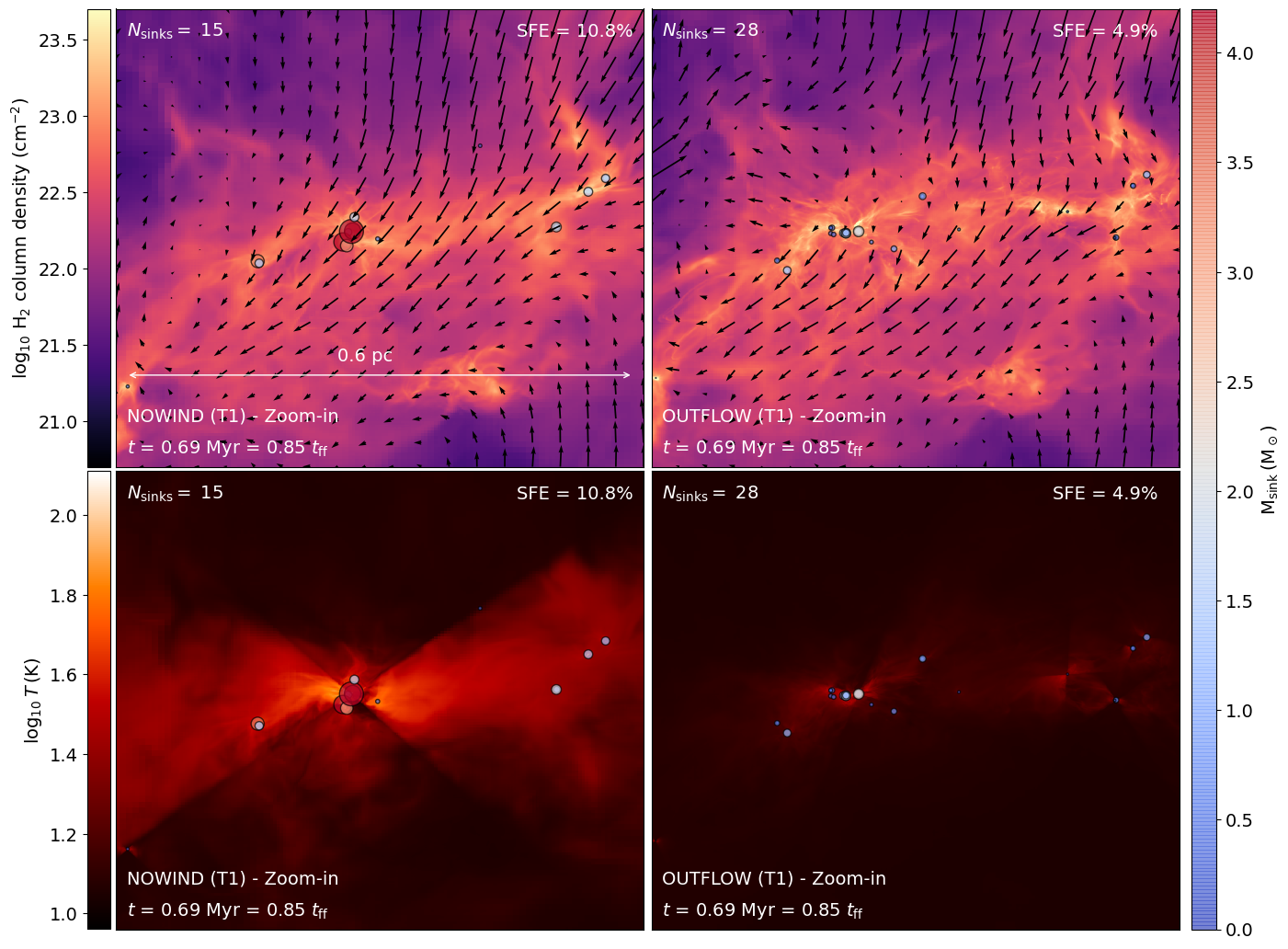}
    \caption{Same as Fig.~\ref{fig:t1_sametime}, but without mass-weighting.}
    \label{fig:t1_sametime_nomw}
\end{figure*}

Fig.~\ref{fig:t1_sametime_nomw} is a non-weighted version of the column density and temperature maps in Fig.~\ref{fig:t1_sametime}. It is evident from comparing both figures that the contrast between the densest regions and the surrounding medium is lower in Fig.~\ref{fig:t1_sametime_nomw}.

\begin{figure*}
    \centering
    \includegraphics[width=\textwidth]{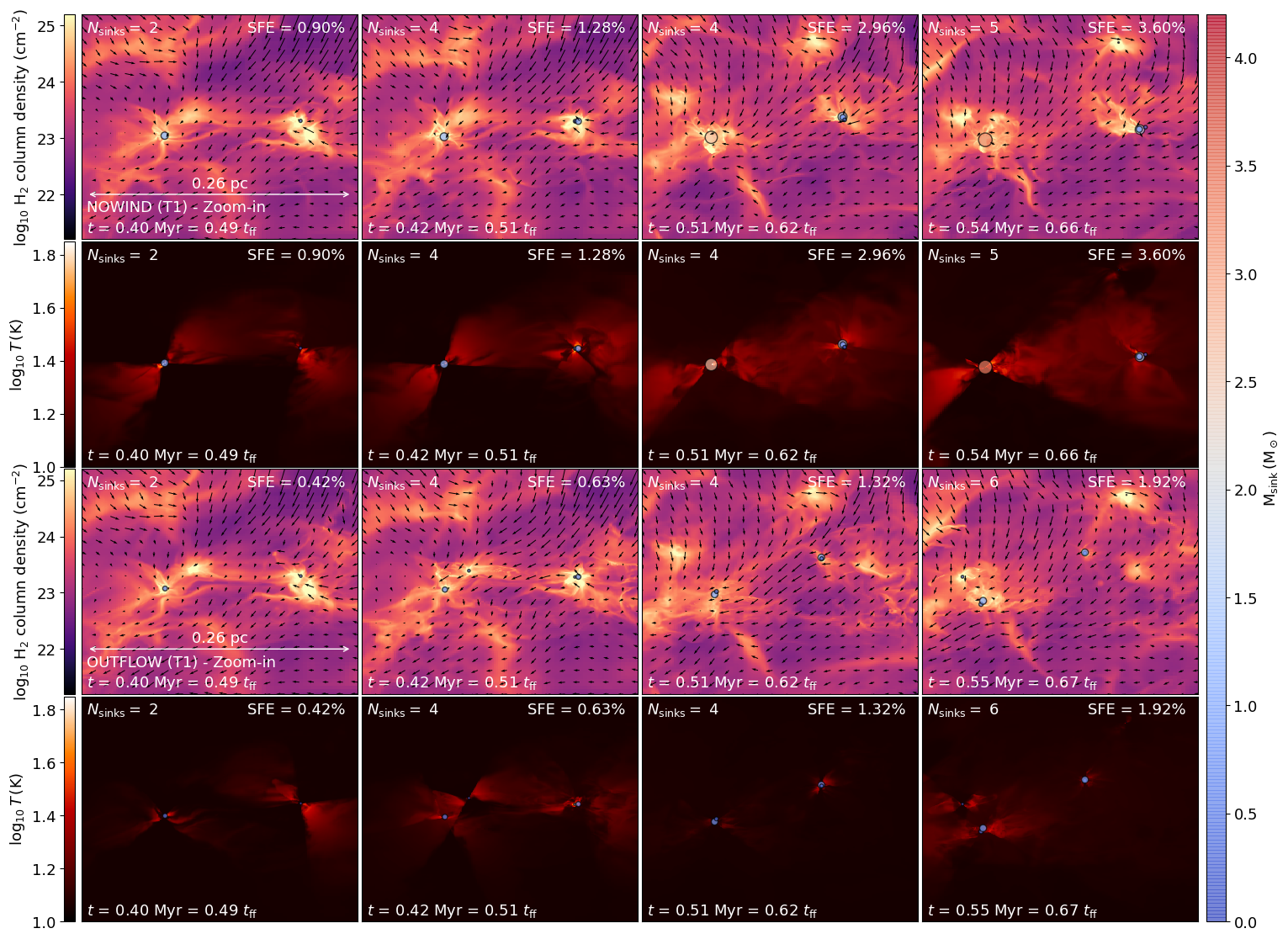}
    \caption{Same as Fig.~\ref{fig:t2_compare}, but for the T1~simulation.}
    \label{fig:t1_compare}
\end{figure*}

Fig.~\ref{fig:t1_compare} is similar to Fig.~\ref{fig:t2_compare}, but for the T1~simulations. This further demonstrates that the additional sink particles in the OUTFLOW simulations form in over-densities created by turbulent flows, which do not collapse in the NOWIND simulations. The figure shows  that at $t=0.40\, \mathrm{Myr}$ (first snapshot), there are two sink particles in both NOWIND and OUTFLOW. Near the sink particle on the left in the OUTFLOW model, a new sink particle forms at $t=0.42\, \mathrm{Myr}$ (second snapshot) and another one forms at $t=0.55\, \mathrm{Myr}$ (last snapshot). The particular over-densities in which these two sinks form also exist in the NOWIND simulation, but there, they do not collapse to form stars.

\begin{figure*}
    \centering
    \includegraphics[width=\textwidth]{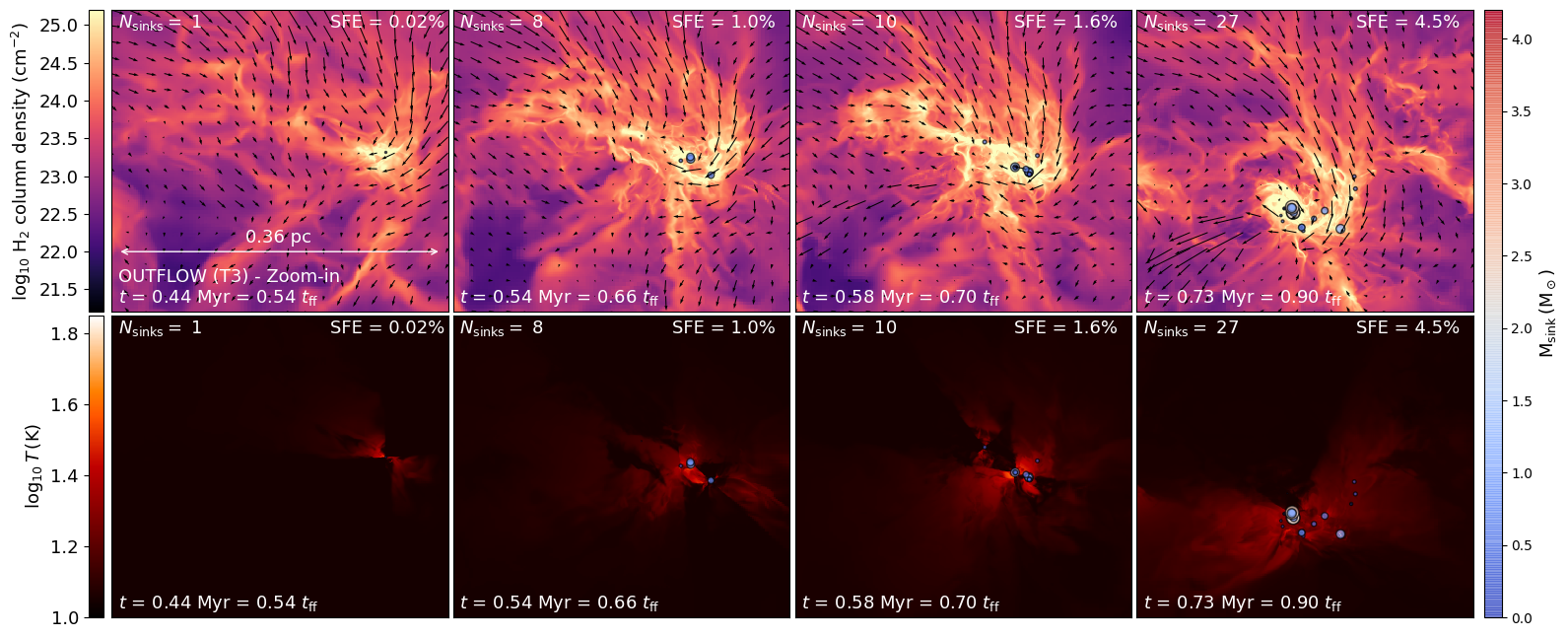}
    \caption{Same as Fig.~\ref{fig:t2_BD}, but for the T3~simulation, showing the formation of sub-stellar objects.}
    \label{fig:t3_BD}
\end{figure*}

\begin{figure*}
    \centering
    \includegraphics[width=\textwidth]{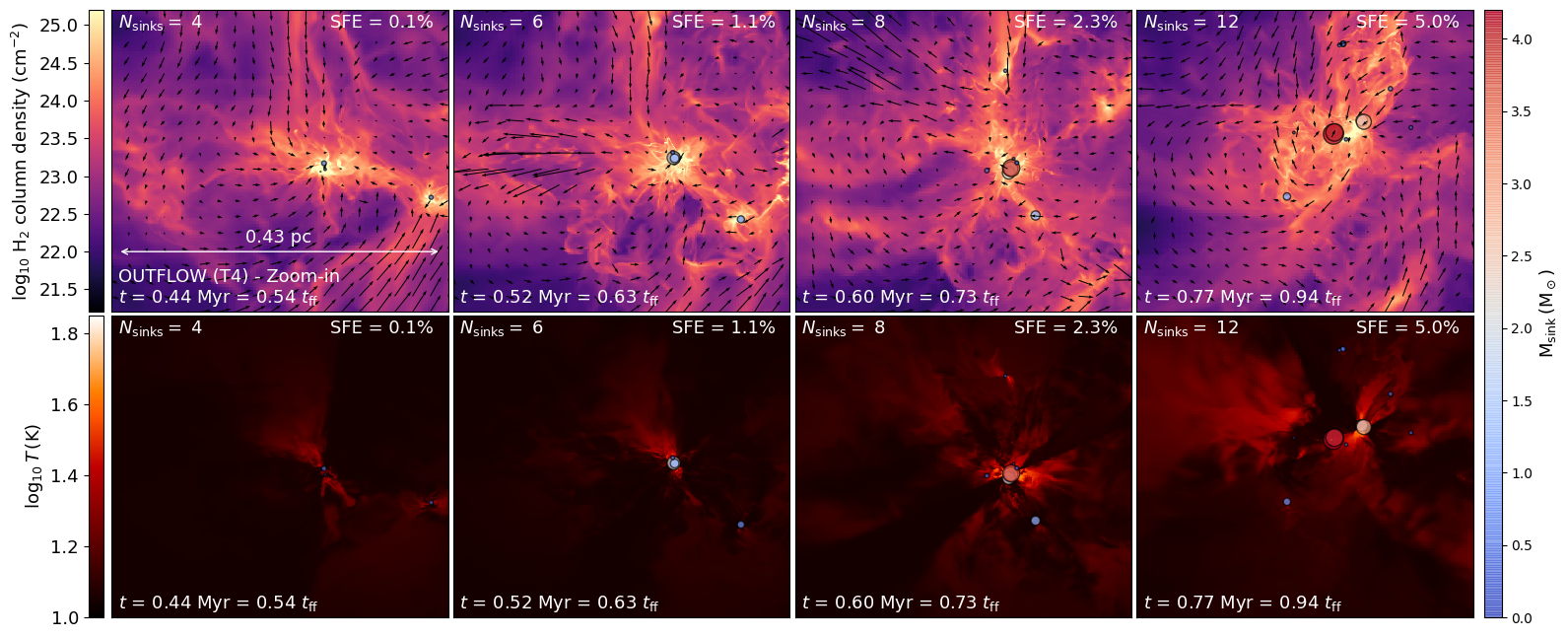}
    \caption{Same as Fig.~\ref{fig:t2_BD}, but for the T4~simulation.}
    \label{fig:t4_BD}
\end{figure*}

Fig.~\ref{fig:t3_BD} and Fig.~\ref{fig:t4_BD} present the evolution of a clump and the formation of VLM objects in the T3 and T4~simulations, respectively.

\bsp	
\label{lastpage}
\end{document}